\documentclass[twocolumn]{aastex7}
\usepackage{amsmath}  \usepackage{amssymb}    \usepackage{booktabs}
\usepackage{caption}
\usepackage{color}
\usepackage{epstopdf}
\usepackage{fancyhdr}
\usepackage{footmisc}   \usepackage{graphicx}
\usepackage{lipsum}       
\usepackage[version=4]{mhchem} \usepackage{multirow}
\usepackage{natbib} 
\usepackage{placeins} 
\usepackage{rotating} 
\usepackage{siunitx}
\usepackage{subcaption} \usepackage{txfonts}
\usepackage{textcomp} 
\usepackage{threeparttable}
\usepackage{url}
\usepackage{xcolor}      \def\kms {km\,s$^{-1}$}         

\begin{document}	
	\title{The TOP-SCOPE Survey of Planck Galactic Cold Clumps: Molecular gas properties}
	\author{Yuebin Yang}
	\affiliation{Xinjiang Astronomical Observatory, Chinese Academy of Sciences, Urumqi 830011, PR China}
	\affiliation{University of Chinese Academy of Sciences, Beijing 100080, PR China}
	\email{}
		
	\author[0000-0003-4910-1390]{Jarken Esimbek}
	\affiliation{Xinjiang Astronomical Observatory, Chinese Academy of Sciences, Urumqi 830011, PR China}
	\affiliation{State Key Laboratory of Radio Astronomy and Technology, A20 Datun Road, Chaoyang District, Beijing, 100101, P. R. China}
	\affiliation{Xinjiang Key Laboratory of Radio Astrophysics, Urumqi 830011, PR China}
	\email[show]{Jarken@xao.ac.cn}
	\correspondingauthor{Jarken Esimbek}
	
	\author[0000-0002-5286-2564]{Tie Liu}	
	\affiliation{Shanghai Astronomical Observatory, Chinese Academy of Sciences, Shanghai 200030, People’s Republic of China}
	\email[show]{liutie@shao.ac.cn} \correspondingauthor{Tie Liu}
	
	\author[0000-0003-3389-6838]{Willem Baan}
	\affiliation{Xinjiang Astronomical Observatory, Chinese Academy of Sciences, Urumqi 830011, PR China}
	\affiliation{Netherlands Institute for Radio Astronomy, ASTRON, 7991 PD Dwingeloo, The Netherlands}
	\email{} 

	\author[0000-0001-8315-4248]{Xunchuan Liu}
	\affiliation{Shanghai Astronomical Observatory, Chinese Academy of Sciences, Shanghai 200030, People’s Republic of China}
	\affiliation{Leiden Observatory, Leiden University, P.O. Box 9513, 2300RA Leiden, The Netherlands}	
	\email{}	
	\author[0000-0003-2412-7092]{Kee-Tae Kim}
	\affiliation{Korea Astronomy and Space Science Institute, 776, Daedeok-daero, Yuseong-gu, Daejeon 34055, Republic of Korea}
	\affiliation{University of Science and Technology, Korea (UST), 217 Gajeong-ro, Yuseong-gu, Daejeon 34113, Republic of Korea}
	\email{}
	
	\author[0000-0003-0933-7112]{Gang Wu}
	\affiliation{Xinjiang Astronomical Observatory, Chinese Academy of Sciences, Urumqi 830011, PR China}
	\affiliation{University of Chinese Academy of Sciences, Beijing 100080, PR China}
	\affiliation{Xinjiang Key Laboratory of Radio Astrophysics, Urumqi 830011, PR China}
	\email{}
	
	\author[0000-0002-4154-4309]{Xindi Tang}
	\affiliation{Xinjiang Astronomical Observatory, Chinese Academy of Sciences, Urumqi 830011, PR China}
	\affiliation{University of Chinese Academy of Sciences, Beijing 100080, PR China}
	\affiliation{State Key Laboratory of Radio Astronomy and Technology, A20 Datun Road, Chaoyang District, Beijing, 100101, P. R. China}
	\affiliation{Xinjiang Key Laboratory of Radio Astrophysics, Urumqi 830011, PR China}
	\email{}
	
	\author[0000-0003-0356-818X]{Jianjun Zhou}
	\affiliation{Xinjiang Astronomical Observatory, Chinese Academy of Sciences, Urumqi 830011, PR China}
	\affiliation{State Key Laboratory of Radio Astronomy and Technology, A20 Datun Road, Chaoyang District, Beijing, 100101, P. R. China}
	\affiliation{Xinjiang Key Laboratory of Radio Astrophysics, Urumqi 830011, PR China}
	\email{}
	
	\author[0000-0001-5494-6238]{Dalei Li}
	\affiliation{Xinjiang Astronomical Observatory, Chinese Academy of Sciences, Urumqi 830011, PR China}
	\affiliation{University of Chinese Academy of Sciences, Beijing 100080, PR China}
	\affiliation{Xinjiang Key Laboratory of Radio Astrophysics, Urumqi 830011, PR China}
	\email{}
	
	\author[0000-0002-8760-8988]{Yuxin He}
	\affiliation{Xinjiang Astronomical Observatory, Chinese Academy of Sciences, Urumqi 830011, PR China}
	\affiliation{University of Chinese Academy of Sciences, Beijing 100080, PR China}
	\affiliation{Xinjiang Key Laboratory of Radio Astrophysics, Urumqi 830011, PR China}
	\email{}
	
	\author{Sung-ju Kang}
	\affiliation{Korea Astronomy and Space Science Institute, 776, Daedeok-daero, Yuseong-gu, Daejeon 34055, Republic of Korea}
	\affiliation{Morescience Inc., 95, Banpo-daero 24-gil, Seocho-gu, Seoul 06526, Republic of Korea}
	\email{}
	
	\author[0000-0002-0776-0753]{Yingxiu Ma}
	\affiliation{Xinjiang Astronomical Observatory, Chinese Academy of Sciences, Urumqi 830011, PR China}
	\affiliation{Xinjiang Key Laboratory of Radio Astrophysics, Urumqi 830011, PR China}
	\email{}
	
	\author{Dongdong Zhou}
	\affiliation{Xinjiang Astronomical Observatory, Chinese Academy of Sciences, Urumqi 830011, PR China}
	\affiliation{Xinjiang Key Laboratory of Radio Astrophysics, Urumqi 830011, PR China}
	\email{}
	
	\begin{abstract}
		We surveyed 2008 Planck Galactic Cold Clumps (PGCCs) in $^{12}\mathrm{CO}$ and $^{13}\mathrm{CO}$ $J=1$--0 lines using the Taeduk Radio Astronomy Observatory (TRAO) 14 m telescope's multi-beam receiver. We detected 2784 ($^{12}\mathrm{CO}$) and 2291 ($^{13}\mathrm{CO}$) velocity components, their closely correlated centroid velocities suggest that $^{12}$CO and $^{13}$CO generally trace kinematically associated gas. PGCCs have low excitation temperatures (mean $\sim$10 K), mean $^{13}\mathrm{CO}$ optical depth $\sim$0.5, and mean $^{13}\mathrm{CO}$-derived H$_2$ column density $4.3\times10^{21}$~cm$^{-2}$. Gas--dust correlations are moderate, with $N_{^{13}\mathrm{CO}}$ more tightly correlated with the dust-derived H$_2$ column density from the PGCC catalog than $I_{^{12}\mathrm{CO}}$. Colder PGCCs tend to have higher CO-to-H$_2$ conversion factor ($X_{\mathrm{CO}}$) and $[\mathrm{H_{2}}]/[^{13}\mathrm{CO}]$ ratio. $X_{\mathrm{CO}}$ increases clearly with the dust-derived H$_2$ column density, consistent with enhanced CO freeze-out in high-column-density gas. Supersonic non-thermal motions are widespread: the Mach number derived from $^{13}\mathrm{CO}$ has a mean of 4.3 and a median of 3.6, increasing slightly with dust-derived H$_2$ column density. Overall, PGCCs are cold but dynamically active, serving as a valuable laboratory for studying the initial conditions of star formation.
	\end{abstract}
	
	\keywords{surveys -- radio lines: ISM -- ISM: molecules -- stars: formation}
	
	\section{Introduction}\label{sec:introdution}
	\setcounter{footnote}{0}
	
	\begin{figure*}[htbp]
		\centering
		\includegraphics[width=0.8\linewidth]{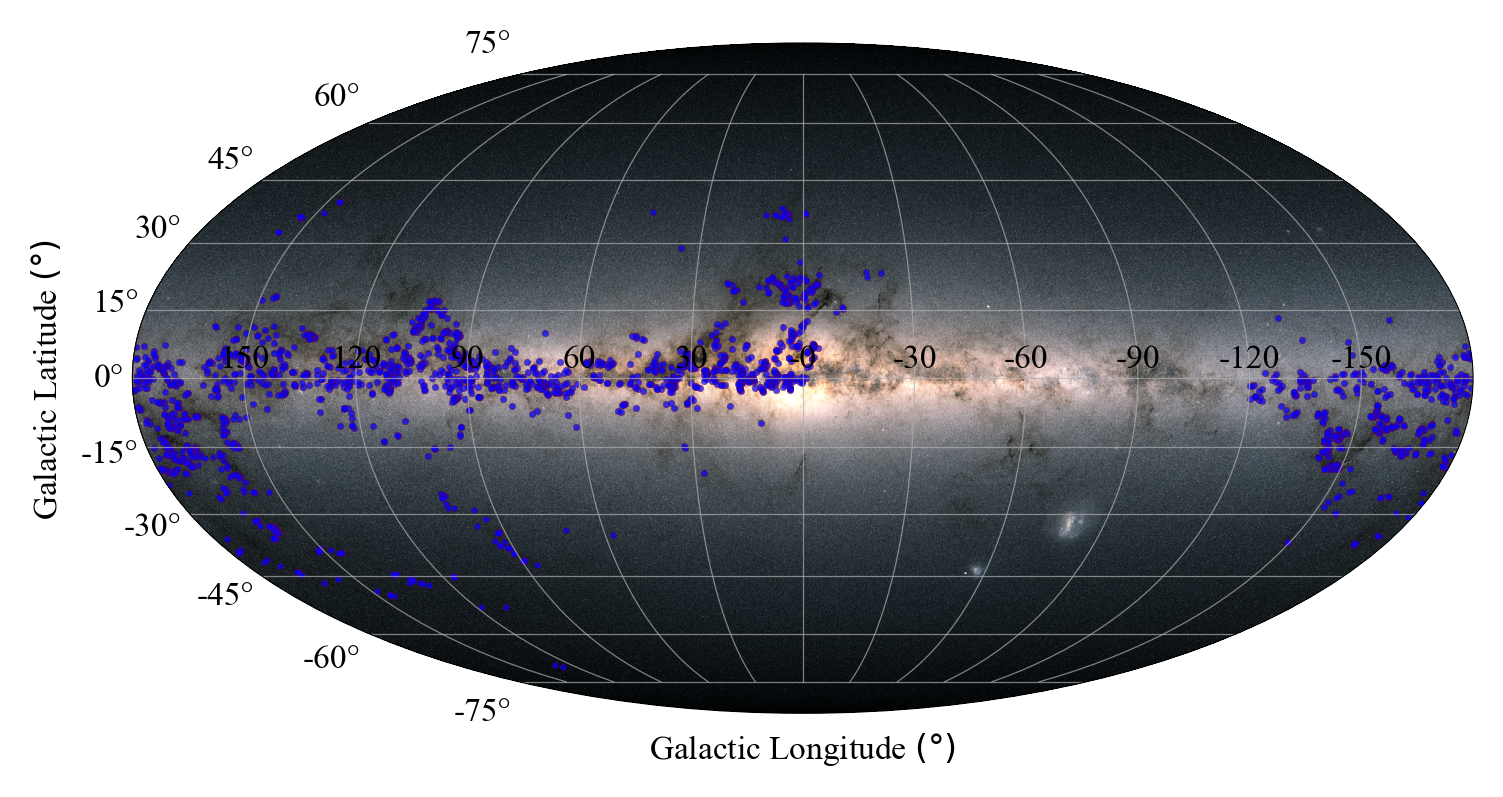}
		\caption{Spatial distribution of the Planck Galactic Cold Clumps (PGCCs) catalog ($N = 2008$ sources, shown as blue points). The background displays the \textit{Gaia} all-sky map combining $G$, $G_{\mathrm{BP}}$, and $G_{\mathrm{RP}}$ bands. The visualization methodology was implemented using Python code adapted from \citet{lagman2022cluster}.}
		\label{fig:1}
	\end{figure*}
	
	Star formation occurs within the cold, dense environments of molecular clouds. Recent advances in observational capabilities have revealed the ubiquitous presence of filamentary structures within these clouds, significantly advancing our understanding of star formation processes \citep{2010A&A...518L.102A,2014prpl.conf...27A}. In particular, massive star formation appears to be intimately connected with these filamentary environments \citep{2018ARA&A..56...41M}. Concurrently, substantial progress has been made in observational studies of dense molecular clumps \citep{2013MNRAS.431.1752U,2018MNRAS.473.1059U,2018A&A...609A.125W}. However, most of these investigated clumps already exhibit signs of ongoing star formation activities, thereby limiting our understanding of their initial properties.
	
	To overcome this limitation, identifying and characterizing prestellar cores---gravitationally bound structures preceding protostellar collapse---is essential \citep{1979ApJS...41..513M,1994MNRAS.268..276W}. Such studies require systematic investigations of cold, dense molecular clumps in their earliest evolutionary stages. The \textit{Planck} satellite's all-sky submillimeter survey, through the Planck Galactic Cold Clumps (PGCCs) catalog, provides an ideal sample for this purpose \citep{2016A&A...594A..28P}. The catalog contains 13,188 sources with dust temperatures typically in the range 13--14.5\,K and H$_2$ column densities spanning $\sim10^{18}$--$10^{23}$\,cm$^{-2}$ across the full catalog, with typical values around $\sim10^{20}$--$10^{21}$\,cm$^{-2}$. Distributed across diverse Galactic environments, these clumps offer an excellent laboratory for probing the initial conditions of star formation.
	
	Early systematic CO molecular-line observations of PGCCs were pioneered by \citet{2012ApJ...756...76W} and \citet{2013ApJ...775L...2L}, who conducted single-point and mapping studies of 674 Planck Galactic Cold Clumps. Their work established the basic physical properties of these cold sources and suggested that the CO abundance may serve as a useful evolutionary indicator, demonstrating the value of PGCCs for studies of the early stages of star formation. Subsequent CO follow-up observations \citep[e.g.,][]{2013ApJS..209...37M,2020ApJS..247...29Z,2021ApJ...920..103X} have further strengthened the role of PGCCs as probes of early star formation. While additional dense molecular tracers have also been used to search for signs of star-forming activity in PGCCs, previous studies have generally remained limited to relatively small samples, restricting a broader statistical characterization of these objects \citep[e.g.,][]{2016ApJ...820...37Y,2019A&A...622A..32L,2024A&A...684A.144B}.
	
	The TOP-SCOPE project is a multi-faceted observational campaign dedicated to the statistical study of Planck Galactic Cold Clumps \citep{2018ApJS..234...28L}. As part of this effort, the present study analyzes 2008 PGCCs from its cornerstone program, the Taeduk Radio Astronomy Observatory (TRAO\footnote{\textcolor{blue}{\url{https://trao.kasi.re.kr/main.php}}}) Key Science Project ``TOP'', nearly tripling the 674-source sample of the pioneering CO survey by \citet{2012ApJ...756...76W}. The TOP survey completed comprehensive $^{12}\mathrm{CO}$ and $^{13}\mathrm{CO}$ ($J=1\text{--}0$) observations of these sources.	
	To avoid duplicating regions already covered by the JCMT Galactic Plane Survey and Gould Belt Survey, the TOP target list excluded those areas, where extensive molecular-line and dust-continuum data are already available \citep{2015MNRAS.453.4264M,2007PASP..119..855W}. As a result, the present sample is not a strictly uniform subsample of the full PGCC catalog, and possible environmental selection effects should be borne in mind when interpreting the statistical results. 
	
	\begin{figure*}[htbp]
		\centering
		\includegraphics[width=\linewidth]{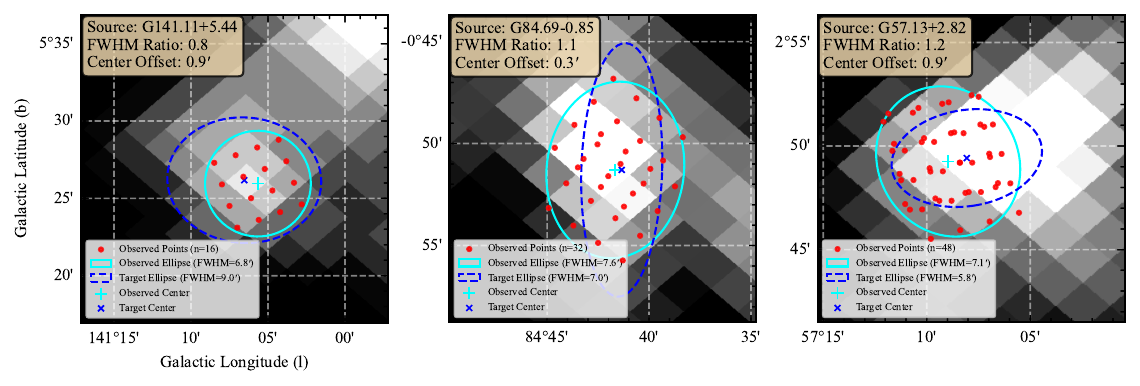}
		\caption{Representative TRAO $^{13}$CO sampled footprints overlaid on the Planck 857 GHz continuum maps. The panels show, from left to right, PGCC G141.11+5.44, G84.69$-$0.85 and G57.13+2.82, illustrating different numbers of sampled beam positions. Red points mark the TRAO beam positions. 
		The cyan ellipse shows the effective sampling footprint fitted to these positions, and the blue dashed ellipse denotes the corresponding PGCC catalog ellipse. The cyan and blue crosses indicate the fitted sampling center and the catalog center, respectively. The fitted-to-catalog FWHM ratio and center offset are given in each panel.}
		\label{fig:ellipse_example}
	\end{figure*}
	
	\begin{figure}[htbp]
		\centering
		\includegraphics[width=\linewidth]{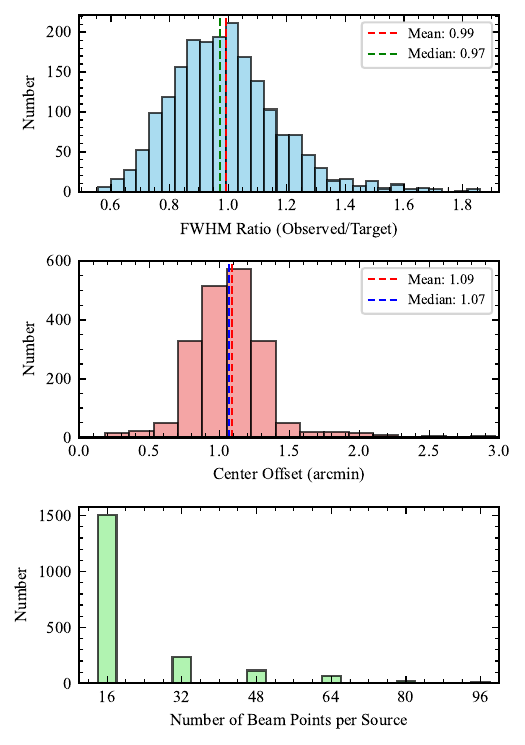}
		\caption{{Statistics of the TRAO sampled footprints relative to the PGCC catalog footprints. The top panel presents the ratio of the effective FWHMs of the fitted sampling ellipse to the PGCC catalog FWHMs, and the middle panel presents the offset between the fitted sampling center and the catalog center. The dashed lines mark the mean and median values. The bottom panel presents the distribution of sampled beam positions per source; 16 positions correspond to one nominal SEQUOIA-TRAO pointing, whereas larger numbers indicate additional offset pointings.}
		}
		\label{fig:footprint_statistics}
	\end{figure}
	
    \section{Observations and Data reduction}\label{sec:observation}   
    
    Figure\,\ref{fig:1} shows the spatial distribution of the 2008 PGCCs observed between December 2015 and March 2018 across the Milky Way\footnote{\textcolor{blue}{\url{https://www.esa.int/ESA_Multimedia/Images/2018/04/Gaia_s_sky_in_colour}}}. The sample covers nearly the full range of Galactic longitude ($0^{\circ}$--$360^{\circ}$), with the exception of an unsurveyed region between $240^{\circ}$ and $338^{\circ}$. Among them, 132 sources (6.6\% of the sample) are located at high Galactic latitudes ($|b| > 30^{\circ}$).
    
    Observations of the $^{12}$CO and $^{13}$CO (1--0) lines toward the PGCCs were conducted using the 14-m Taeduk Radio Astronomy Observatory (TRAO) telescope, equipped with the SEQUOIA-TRAO receiver \citep{2019JKAS...52..227J}. This system employs a 4$\times$4 array (16 pixels) operating at 85--115 GHz, yielding receiver noise temperatures of 50--80 K and system temperatures of 150--450 K across the band. The two lines were observed simultaneously. The backend FFT spectrometer provided a velocity resolution better than 0.1 km s$^{-1}$ and a bandwidth of 60 MHz ($\sim$160~\kms\ at 115 GHz). The same receiver and backend configuration were used throughout the observing campaigns. Examination of the spectra reveals no significant absorption features caused by emission at reference positions. Flux calibration was monitored through regular observations of the standard spectral-line source IRC\,10216. The HPBWs are approximately $47^{\prime\prime}$ for $^{12}$CO and $49^{\prime\prime}$ for $^{13}$CO. Main-beam efficiencies ($\eta_{\mathrm{mb}}$) were adopted separately for each observing year: $\eta_{\mathrm{mb}}=46\%$ for $^{12}$CO and $49\%$ for $^{13}$CO in 2016, $41\%$ and $46\%$ in 2017. For the 2018 data, we adopted the 2019 efficiency values recommended by the observatory, namely $43.1\%$ for $^{12}$CO and $48.1\%$ for $^{13}$CO, because efficiency measurements were not conducted in 2018.
    
    {The observations analyzed in this paper were obtained in position-switched mode.
    For each target, the on-source and off-source integration times were 30~s each, and a nearby reference position was used as the off position for baseline and sky subtraction. Each nominal SEQUOIA-TRAO pointing samples up to 16 discrete beam positions with an inter-beam spacing of approximately twice the HPBW. Most PGCCs were observed with one nominal pointing, while a subset was observed with additional offset pointings, resulting in larger numbers of sampled beam positions, such as 32, 48, or 64. Thus, the data provide sparse, discrete spatial sampling of each target rather than fully sampled or Nyquist-sampled maps.}
    
    {Figure\,\ref{fig:ellipse_example} shows representative examples of the sampled footprints, and Figure\,\ref{fig:footprint_statistics} summarizes the corresponding full-sample statistics. The mean FWHM of the fitted sampling footprints is $(6.9 \pm 0.2)^\prime$, comparable to the mean PGCC catalog FWHM of $(7.1 \pm 1.3)^\prime$, and the mean offset between the fitted sampling centers and the catalog centers is $1.0^\prime$. For 92\% of the sources, all sampled positions lie within the corresponding catalog ellipse, for the remaining $\sim$150 sources, more than 80\% of the sampled positions are still enclosed by the catalog ellipse. These footprint statistics indicate that the TRAO pointings generally sample the intended catalog regions, but they should not be interpreted as measurements of the intrinsic CO-emitting extent or as evidence for a full one-to-one spatial correspondence between the CO emission and the Planck dust emission.}
    
    Data reduction and spectral analysis were performed using the \texttt{GILDAS}\footnote{\textcolor{blue}{\url{https://www.iram.fr/IRAMFR/GILDAS/gildasli2.html}}} software package, specifically the \texttt{CLASS} and \texttt{GREG} modules for Gaussian fitting and line-profile visualization \citep{2000ASPC..217..299G,2005sf2a.conf..721P}. 
    For each PGCC, the source-averaged spectrum used in the statistical analysis was constructed as the arithmetic mean of the spectra from all sampled beam positions after baseline subtraction and conversion to main-beam temperature. 
    {The rms noise level, $\sigma_{\rm rms}$, was estimated from line-free velocity channels. The velocity interval of the detected emission was first identified from the source-averaged spectrum. Individual beam positions were then classified as detections when the peak line emission within this velocity interval satisfied a signal-to-noise ratio (S/N) greater than 3. Single-peaked profiles with approximately Gaussian shapes were fitted with one component, whereas spectra with multiple significant peaks were fitted with multiple Gaussian components.}
    
    {For ambiguous or blended profiles, which occurred primarily in $^{12}$CO, the corresponding source-averaged $^{13}$CO spectrum was used as an auxiliary guide for assigning the number of Gaussian velocity components. 
    When the $^{13}$CO spectrum showed one significant velocity component, the associated $^{12}$CO emission was fitted with a single Gaussian component. 
    When multiple $^{13}$CO components were detected, the $^{12}$CO profile was fitted with the corresponding number of components, using the $^{13}$CO velocity structure as guidance. After fitting, the results were examined source by source by comparing the fitted $^{12}$CO and $^{13}$CO centroid velocities. Sources exhibiting unusually large centroid discrepancies were re-examined and refitted when necessary, in order to reduce possible biases introduced by manual decomposition. No automated AIC/BIC model-selection 
    procedure was adopted, the statistical detection criterion was the $3\sigma$ threshold, supplemented by the observed line-profile morphology and consistency between the two isotopologues.}
    
    {Multiple velocity components toward a given PGCC were treated as line-of-sight components and were not assumed to be physically connected. 
    For comparisons with the source-level quantities provided in the PGCC catalog, additive quantities, such as integrated intensity and column density, were summed over all identified components, whereas non-additive quantities, such as excitation temperature, velocity dispersion, and Mach number, were represented by integrated-intensity-weighted averages. 
    The weights were defined as $w_i = I_i/\sum_j I_j$, where $I_i$ is the integrated intensity of the $i$th component. 
    This procedure provides a source-averaged summary of the molecular emission detected along each line of sight and should not be interpreted as a complete representation of the intrinsic three-dimensional kinematic structure of the corresponding Planck clump.}
      
    The dust-derived H\textsubscript{2} column density ($N_{\mathrm{H}_{2}\mathrm{-dust}}$) and dust temperature ($T_{\mathrm{dust}}$) values were obtained directly from the \textit{Planck} Galactic Cold Clumps catalog, available through the Planck Legacy Archive on Vizier \citep{https://doi.org/10.26093/cds/vizier.35940028}. For the main analysis in this paper, we adopt the catalog values derived with a fixed dust emissivity index of $\beta=2$, using the catalog H$_2$ column-density entry $N_{\mathrm{H}_2}$ and its corresponding uncertainty entry, in order to maintain a uniform definition of the dust parameters and to reduce spurious correlations introduced by free-$\beta$ modified-blackbody fitting \citep{2009ApJ...696..676S,2013A&A...556A..63J}. At the \textit{Planck} resolution, $N_{\mathrm{H_2-dust}}$ may be systematically underestimated because colder and denser substructures are not fully resolved within the parent clump \citep{2016A&A...594A..22P}. In our sample, $N_{\mathrm{H_2-dust}}$ spans $2.0\times10^{20}$--$5.9\times10^{22}~\mathrm{cm}^{-2}$, with a mean of $2.1\times10^{21}~\mathrm{cm}^{-2}$ and a median of $1.3\times10^{21}~\mathrm{cm}^{-2}$. The typical $1\sigma$ relative uncertainty is $\sim$50\%. The $T_{\mathrm{dust}}$ values range from 9 to 30 K, with a median of 13 K, and the corresponding $1\sigma$ statistical uncertainty has a median value of 1.0 K across the sample.
    
    \begin{figure*}[htbp]
    	\centering
    	\includegraphics[width=0.95\linewidth]{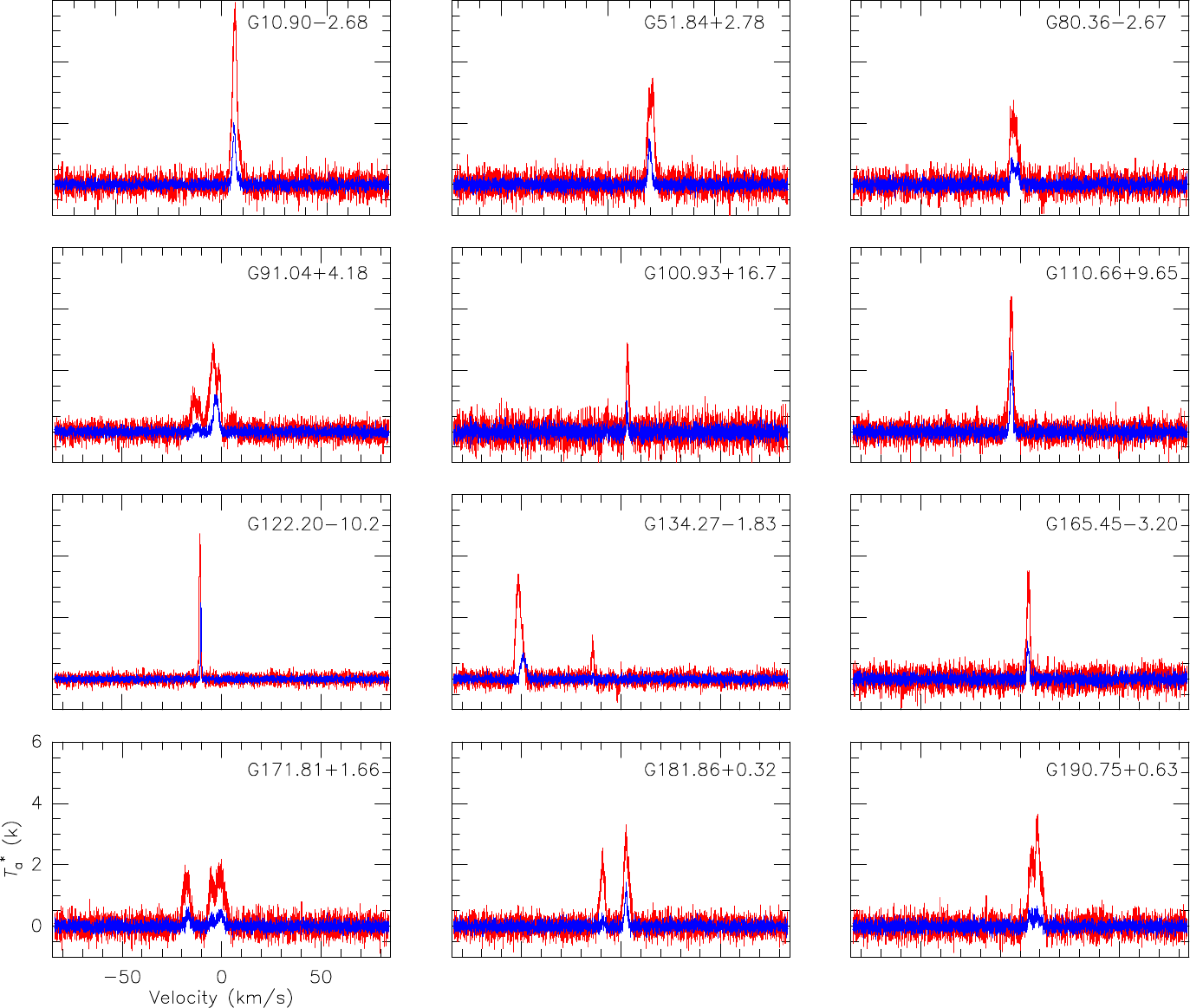}	
    	\caption{$^{12}$CO and $^{13}$CO ($J = 1\text{–}0$) spectral line emissions toward example Planck Galactic Cold Clumps (PGCCs). The $^{12}$CO and $^{13}$CO lines are shown in red and blue, respectively.}
    	\label{fig:spe}
    \end{figure*}
      
    \section{Results}\label{sec:results}        
	\subsection{Observed line parameters} 
	
	\begin{figure*}[htbp]
	   	\centering
	   	\includegraphics[width=0.9\linewidth]{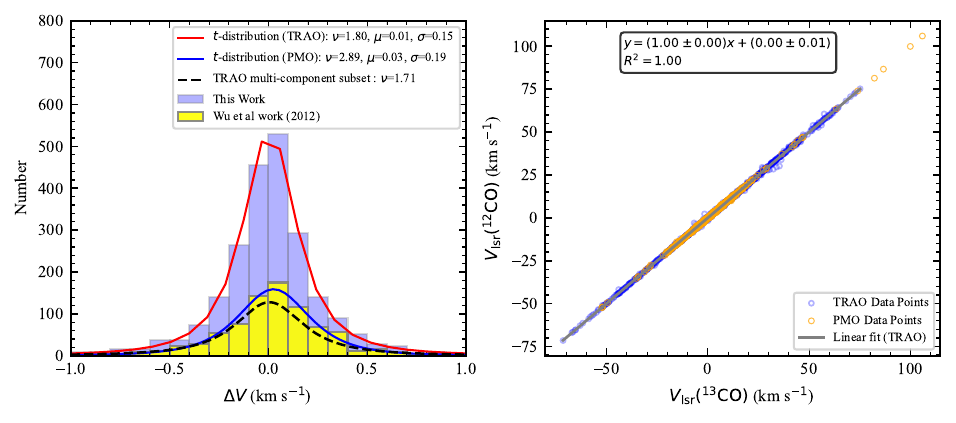}
	   	\caption{{Left panel: distribution of the centroid-velocity difference 
	   	$\Delta V$. Blue and yellow histograms represent the new TRAO sample and the PMO sample from \citet{2012ApJ...756...76W}, respectively. Solid curves show the best-fit Student's $t$-distributions for the two full samples, with the corresponding fit parameters indicated. The dashed curve shows the fit for the TRAO subset associated with PGCCs exhibiting multiple detected $^{13}$CO  velocity components. Right panel: comparison of the $^{12}$CO and $^{13}$CO centroid velocities for the same matched components. 
	   	The solid line shows the linear fit to the TRAO data.}}
	   	\label{fig:velocity}
	\end{figure*}
	
	\begin{figure*}[htbp]
	   	\includegraphics[width=\linewidth]{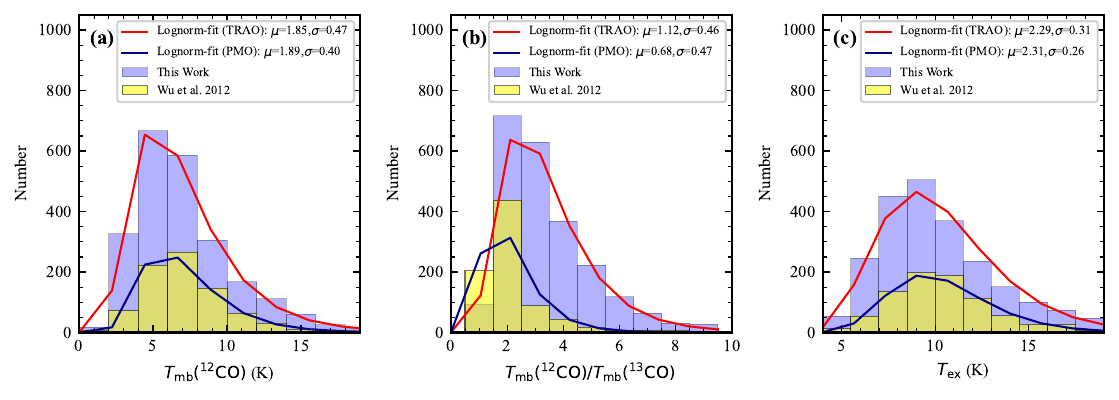}
	   	\caption{Distributions and lognormal fits for three parameters derived from the $^{12}\mathrm{CO}$ and $^{13}\mathrm{CO}$ observations. (a) $T_{\mathrm{mb}}(^{12}\mathrm{CO})$, the main-beam brightness temperature of $^{12}$CO. (b) The ratio $T_{\mathrm{mb}}(^{12}\mathrm{CO})/T_{\mathrm{mb}}(^{13}\mathrm{CO})$. (c) Excitation temperature ($T_{\mathrm{ex}}$). Histograms show data from the TRAO telescope (this work) and comparative data from the PMO telescope \citep{2012ApJ...756...76W}. Solid curves represent the best-fit lognormal distributions, with the fitted parameters ($\mu$, $\sigma$) indicated in each panel.}
	   	\label{fig:Tmb}
	\end{figure*}
	
	\begin{figure}[htbp]
		\centering
		\includegraphics[width=\linewidth]{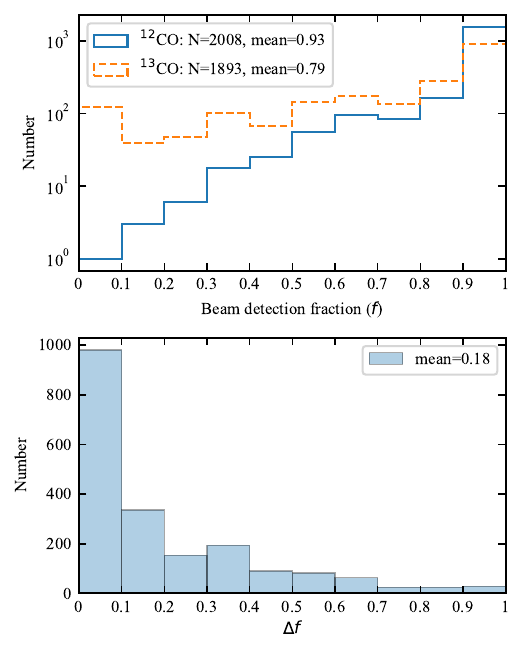}
		\caption{{Distribution of the beam detection fractions for the $^{12}$CO and $^{13}$CO ($J=1$--0) lines (upper panel), and of the detection-fraction difference $\Delta f=f_{12}-f_{13}$ (bottom panel). For each PGCC, $f_{\rm det}=N_{\rm det}/N_{\rm beam}$, where $N_{\rm beam}$ is the total number of sampled beam positions associated with the source and $N_{\rm det}$ is the number of beam positions with detected line emission.}
		}		
		\label{fig:detection_fraction}
	\end{figure}
	
	The observed line parameters analyzed in this section are derived from Gaussian fitting to the source-averaged  spectra described in Section~\ref{sec:observation}. Figure~\ref{fig:spe} presents representative examples of these averaged $^{12}$CO and $^{13}$CO spectra. The main-beam brightness temperature is defined as
	
	\begin{equation}
	   	T_{\mathrm{mb}}=\frac{T_{\mathrm{a}}^{*}}{\eta_{\mathrm{mb}}},
	\end{equation}    
	
	\noindent where $T_{\mathrm{a}}^{*}$ is the antenna temperature and $\eta_{\mathrm{mb}}$ is the main-beam efficiency appropriate for the observing year. All the resulting line parameters are listed in Table~\ref{tab:line_parameters} (Appendix~A).
	
	The analysis yielded 2784 $^{12}$CO and 2291 $^{13}$CO velocity components from Gaussian fitting of all detected spectral features. $^{12}$CO emission is detected toward all 2008 sources, and $^{13}$CO is detected toward 1895 of them. A total of 561 PGCCs (28\%) exhibit at least two $^{12}$CO velocity components, of which 151 contain more than two. Among the 1895 sources with detected $^{13}$CO emission, 302 (16\%) show multiple $^{13}$CO velocity components.
	
	{We performed a direct statistical comparison between the $^{12}$CO and $^{13}$CO centroid velocities relative to the Local Standard of Rest (LSR) by examining the distribution of their velocity differences, defined as \(\Delta V = V_{\mathrm{lsr}}(^{12}\mathrm{CO}) - V_{\mathrm{lsr}}(^{13}\mathrm{CO})\). As shown in Figure~\ref{fig:velocity}, the $\Delta V$ distribution is sharply peaked near 0~\kms with extended wings, and is therefore empirically described by a Student's $t$-distribution rather than a single Gaussian profile. 
	The full TRAO sample gives a fitted value of $\nu \approx 1.8$, compared with $\nu \approx 2.9$ for the comparison sample of \citet{2012ApJ...756...76W} analyzed in the same way. Smaller values of $\nu$ correspond to heavier tails and therefore stronger departures from Gaussianity. In the limit $\nu \rightarrow \infty$, the $t$-distribution converges to a Gaussian form.}
	
	{As an additional check, we fitted the $\Delta V$ distribution for components belonging to PGCCs with multiple detected velocity components in order to examine whether the extended wings are primarily associated with line-of-sight velocity complexity. 
	This subset yields $\nu \approx 1.7$, numerically close to the value obtained for the full TRAO sample. The similarity of these fitted values suggests that the heavy-tailed behavior is not markedly enhanced in the multi-component subset and is therefore unlikely to arise solely from the presence of multiple velocity components along the line of sight. 
	More generally, the extended wings may reflect a combination of effects, including residual uncertainties in the manual decomposition of complex profiles, opacity-related centroid shifts or line-profile asymmetries in the generally more optically thick $^{12}$CO transition, and the small but non-identical beam sizes of the $^{12}$CO and $^{13}$CO observations. These effects cannot be fully separated in the present single-pointing analysis. Nevertheless, the sharp peak near $\Delta V \approx 0$~\kms and the tight centroid-velocity correlation shown in the right panel of Figure~\ref{fig:velocity} indicate that the two isotopologues generally trace kinematically associated gas along the same lines of sight.}
	
	To facilitate direct comparison with \citet{2012ApJ...756...76W}, we fitted the parameter distributions with a lognormal function of the form
	\begin{equation}
	   	f_X(x;\mu,\sigma)=\frac{1}{x\sigma\sqrt{2\pi}}
	   	\exp\left[-\frac{(\ln x-\mu)^2}{2\sigma^2}\right], \quad x>0,
	\end{equation}
	\noindent where $\mu$ and $\sigma$ are the mean and standard deviation of $\ln x$.
	
	The goodness of fit was assessed using the Kolmogorov--Smirnov (K--S) test. Throughout this section, distributions with $p<0.05$ are taken to reject the null hypothesis of a lognormal distribution.
	
	The statistical comparisons in Figure~\ref{fig:Tmb} are restricted to components with detected $^{13}$CO emission. This choice both preserves direct consistency with \citet{2012ApJ...756...76W} and ensures that all compared quantities are defined on the same subset, since the line ratio $T_{\mathrm{mb}}(^{12}\mathrm{CO})/T_{\mathrm{mb}}(^{13}\mathrm{CO})$, the excitation temperature, the $^{13}$CO optical depth, and the subsequent $^{13}$CO-based physical parameters all require detected $^{13}$CO emission.
	
	{For this subset, the mean $T_{\mathrm{mb}}(^{12}\mathrm{CO})$ is 7.1~K and the median is 6.3~K, in close agreement with the values reported by \citet{2012ApJ...756...76W} (mean 7.2~K; median 6.8~K). By contrast, the mean ratio $T_{\mathrm{mb}}(^{12}\mathrm{CO})/T_{\mathrm{mb}}(^{13}\mathrm{CO})$ increases from $\sim 2$ in the PMO sample to $\sim 3$ in the TRAO sample, corresponding to a rise of roughly 50\%. This difference is likely related to the construction of the TRAO source-averaged spectra. For each PGCC, our spectra were averaged over multiple sampled beam positions, whereas \citet{2012ApJ...756...76W} used single-point measurements for sources with detected $^{13}$CO emission. As shown in Figure~\ref{fig:detection_fraction}, $^{12}$CO is detected over a larger fraction of the sampled beam positions than $^{13}$CO. Therefore, averaging over all sampled positions can dilute the source-averaged $^{13}$CO peak temperature more strongly than the $^{12}$CO peak temperature, leading to a larger footprint-averaged line ratio. This larger ratio is not inconsistent with the similar excitation temperatures obtained in the two samples, because $T_{\rm ex}$ is derived from the $^{12}$CO peak brightness under the optically thick assumption, and the characteristic $T_{\rm mb}(^{12}\mathrm{CO})$ values are nearly unchanged. The enhanced $T_{\mathrm{mb}}(^{12}\mathrm{CO})/T_{\mathrm{mb}}(^{13}\mathrm{CO})$ ratio therefore mainly reflects the relatively weaker source-averaged $^{13}$CO emission and should not be interpreted as the intrinsic line ratio at positions where both isotopologues are detected.}
	
	{A statistical comparison of the line widths is shown in Figure~\ref{fig:width}. We adopt the median as a representative value to reduce the influence of outliers. The median FWHMs are 2.5 and 1.6~\kms\ for $^{12}$CO and $^{13}$CO, respectively, compared with 1.8 and 1.1~\kms\ in \citet{2012ApJ...756...76W}, the present values are therefore larger by roughly 40\%--50\%. As discussed above, the TRAO source-averaged spectra were constructed from multiple sampled beam positions across each PGCC footprint. In addition to affecting the source-averaged brightness temperatures, this averaging procedure can broaden the resulting line profiles. Weak peripheral emission and modest centroid-velocity variations among beam positions may contribute additional line-wing emission when averaged together, leading to systematically broader profiles than those obtained from single-point observations. Such effects may partly contribute to the larger line widths measured in the TRAO sample and, consequently, to the higher non-thermal velocity dispersions and Mach numbers derived in the present study, in addition to possible differences in sample selection and physical environment.} 
		
	{Applying the kinematic criterion of \citet{1983ApJ...266..309M}, $\mathrm{FWHM}(^{13}\mathrm{CO})>1.3$~\kms, identifies 63\% of the $^{13}$CO components, corresponding to 60\% of all PGCCs, as candidate high-mass clumps. Even with the stricter threshold of 2.0~\kms\ from \citet{2009A&A...507..369W}, the fractions remain 34\% and 32\%, respectively. Unlike the PMO sample, the $^{13}$CO line-width distribution in the present sample is not consistent with a lognormal form. While the TRAO observations do not provide fully sampled maps, the multi-beam sampling strategy characterizes molecular emission over a larger fraction of each PGCC footprint than a single-point measurement. The resulting source-averaged quantities therefore provide a complementary view of the global molecular properties of PGCCs on the spatial scales sampled by the survey.}

	\subsection{Derived physical parameters}
    
    \begin{figure*}[htbp]
    	\includegraphics[width=\linewidth]{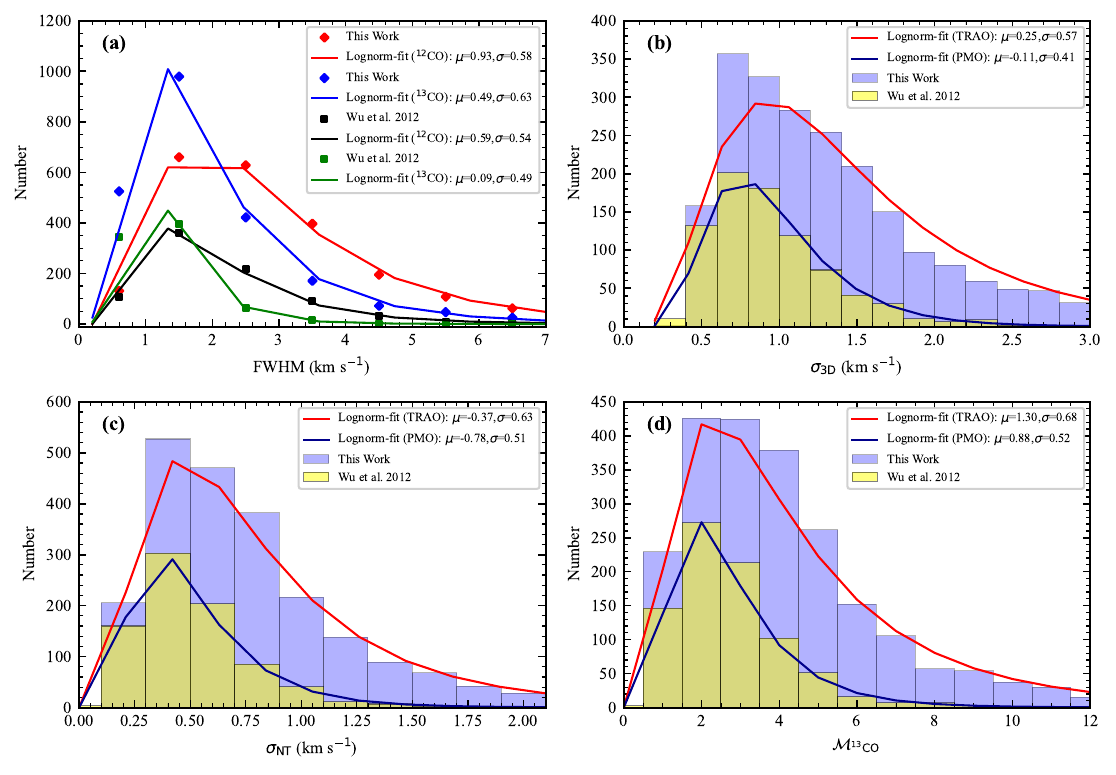}
    	\caption{Distributions of kinematic parameters derived from the $^{12}\mathrm{CO}$ and $^{13}\mathrm{CO}$ observations. (a) Full width at half maximum (FWHM) of the $^{12}\mathrm{CO}$ and $^{13}\mathrm{CO}$ lines. (b) Three-dimensional velocity dispersion ($\sigma_{\mathrm{3D}}$). (c) Non-thermal velocity dispersion ($\sigma_{\mathrm{NT}}$). (d) Mach number, $\mathcal{M}_{^{13}\mathrm{CO}}=\sigma_{\rm NT}/c_s$ Histograms show the TRAO data (this work) and the PMO data from \citet{2012ApJ...756...76W}. Solid curves represent lognormal fits with the best-fit parameters ($\mu$, $\sigma$) indicated.}
    	\label{fig:width}
    \end{figure*}
    
    \begin{figure}[htbp]
    	\includegraphics[width=\linewidth]{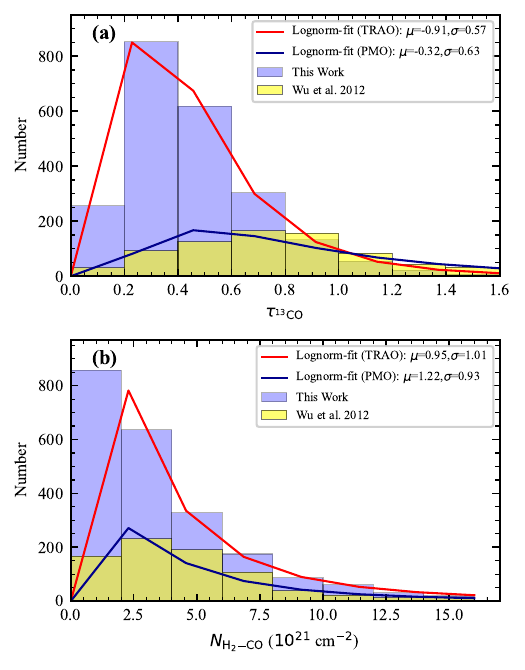}
    	\caption{Distributions of physical parameters derived from the $^{13}\mathrm{CO}$ observations. (a) Optical depth ($\tau_{^{13}\mathrm{CO}}$) of the $^{13}\mathrm{CO}$ line. (b) CO-based H$_2$ column density, $N_{\mathrm{H_2-CO}}$, in units of $10^{21}~\mathrm{cm}^{-2}$. Histograms show the TRAO data (this work) and the PMO data from \citet{2012ApJ...756...76W}. Solid curves represent lognormal fits with the best-fit parameters ($\mu$, $\sigma$) indicated.}
    	\label{fig:op_and_nh}
    \end{figure}
    
    The excitation temperature $T_{\mathrm{ex}}$ and optical depth $\tau$ were derived from the radiative-transfer equation
    
    \begin{equation}
    	\begin{split}
    		T_{\mathrm{mb}} &= \frac{h\nu}{k}
    		\left[\frac{1}{\exp(h\nu/kT_{\mathrm{ex}})-1}
    		-\frac{1}{\exp(h\nu/kT_{\mathrm{bg}})-1}\right] \\
    		&\quad \times [1-\exp(-\tau)] f,
    	\end{split}
    \end{equation}
    where $T_{\mathrm{bg}}=2.73$~K is the cosmic microwave background temperature and $f$ is the beam-filling factor. Assuming optically thick $^{12}\mathrm{CO}$ emission ($\tau_{^{12}\mathrm{CO}}\gg 1$) and $f=1$, the excitation temperature is written as
    
    \begin{equation}
    	T_{\mathrm{ex}}=5.53\left[\ln\left(1+\frac{5.53}{T_{\mathrm{mb}}(^{12}\mathrm{CO})+0.819}\right)\right]^{-1},
    \end{equation}
    
    \noindent following \citet{1998AJ....116..336N} and \citet{rohlfs2013tools}.
    
    The resulting excitation temperature has a mean value of 10.4~K and a median of 9.7~K. The mean is close to the value of 10.1~K reported by \citet{2012ApJ...756...76W}. According to the K--S criterion described above ($p<0.05$), however, the $T_{\rm ex}$ distribution in the present sample is not consistent with a lognormal form.
    
    Under the LTE approximation, $^{12}$CO and $^{13}$CO are taken to share the same excitation temperature. 
    The $^{13}$CO optical depth is then estimated from
    
    \begin{equation}
    	\tau_{^{13}\mathrm{CO}}=
    	-\ln\left[1-\frac{T_{\mathrm{mb}}(^{13}\mathrm{CO})}{T_{\mathrm{mb}}(^{12}\mathrm{CO})}\right].
    \end{equation}
    
    {The derived $^{13}$CO optical-depth distribution has a mean of 0.5 and a median of 0.4, compared with the mean value of 0.9 reported by \citet{2012ApJ...756...76W}. 
    This lower characteristic optical depth is consistent with the larger $T_{\mathrm{mb}}(^{12}\mathrm{CO})/T_{\mathrm{mb}}(^{13}\mathrm{CO})$ ratio found in the present sample, from which $\tau_{^{13}\mathrm{CO}}$ is directly inferred. 
    As discussed above, the difference may partly reflect the broader source-level spatial sampling of the TRAO spectra relative to the central single-point measurements of \citet{2012ApJ...756...76W}, although sample differences may also contribute.} These values indicate that $^{13}$CO is generally optically thin to moderately thick in the present sample. 
    Therefore, the strict limit $\tau_{^{13}\mathrm{CO}}\ll 1$ is not valid for every component, although most sources remain below unity. 
    In the subsequent $N_{^{13}\mathrm{CO}}$ calculation, we do not adopt the strict optically thin limit, instead, we retain the finite-optical-depth correction factor $\tau_{^{13}\mathrm{CO}}/(1-e^{-\tau_{^{13}\mathrm{CO}}})$.
    
    The $N_{^{13}\mathrm{CO}}$ column density was calculated following \citet{1997ApJ...476..781B,1991ApJ...374..540G}:
    
    \begin{equation}
    	N_{^{13}\mathrm{CO}}=
    	2.42\times10^{14}
    	\frac{\tau_{^{13}\mathrm{CO}}}{1-e^{-\tau_{^{13}\mathrm{CO}}}}
    	\frac{1+0.88/T_{\mathrm{ex}}}{1-e^{-5.29/T_{\mathrm{ex}}}}
    	\int T_{\mathrm{mb}}(^{13}\mathrm{CO})\,dv.
    \end{equation}
    
    To enable direct comparison with \citet{2012ApJ...756...76W}, we adopt the same abundance ratio, $[\mathrm{H}_2]/[^{13}\mathrm{CO}] = 8.9\times10^{5}$ \citep{1980ApJ...237....9M}. We emphasize that this assumption is used only to derive the CO-based H$_2$ column density, denoted here as $N_{\mathrm{H_2-CO}}$. We also note that the adopted ratio is a canonical constant for comparison purposes and may vary with environment (see Section~\ref{abundance}). Using this ratio, we obtain $N_{\mathrm{H_2-CO}}$ with a mean of $4.3\times10^{21}~\mathrm{cm}^{-2}$ and a median of $2.7\times10^{21}~\mathrm{cm}^{-2}$, close to the mean value of $4.4\times10^{21}~\mathrm{cm}^{-2}$ reported by \citet{2012ApJ...756...76W}. Figure~\ref{fig:op_and_nh} presents the corresponding histograms of $\tau_{^{13}\mathrm{CO}}$ and $N_{\mathrm{H_2-CO}}$.
    
    The CO-to-H$_2$ conversion factor is defined as
    
    \begin{equation}
    	X_{\rm CO\text{-}to\text{-}H_2}=\frac{N_{\mathrm{H_2-dust}}}{\sum_i I_{\rm CO}^{i}}.
    \end{equation}
    {where $I^i_{^{12}{\rm CO}}$ is the integrated intensity of the $i$th detected $^{12}$CO 
    velocity component. This definition uses the dust-derived H$_2$ column density and is 
    therefore an empirical, source-averaged quantity in the present analysis.}
    
    Assuming that the gas is close to LTE, we take the excitation temperature to approximate the kinetic temperature, $T_{\mathrm{ex}} \approx T_{\mathrm{kin}}$. This approximation is commonly adopted for cold molecular gas and is likely reasonable for much of the PGCCs sample, whose characteristic densities are typically of order $10^{4}$~cm$^{-3}$ \citep{2016A&A...594A..28P}. However, it may break down in lower-density or subthermally excited gas, where the level populations are not fully thermalized \citep{1999ApJ...517..209G,2015PASP..127..299S}. In addition, observational effects such as beam dilution, temperature gradients, or self-absorption may further complicate the interpretation of $T_{\mathrm{ex}}$.
    
    The non-thermal velocity dispersion of $^{13}$CO is then given by  
    \begin{equation}
    	\sigma_{\mathrm{NT}}=
    	\left[\sigma_{^{13}\mathrm{CO}}^2-\frac{k_{\mathrm{B}}T_{\mathrm{ex}}}{m_{^{13}\mathrm{CO}}}\right]^{1/2},
    \end{equation}    
    where $\sigma_{^{13}\mathrm{CO}}=\Delta V(^{13}\mathrm{CO})/\sqrt{8\ln 2}$ is the observed $^{13}$CO velocity dispersion, $k_{\mathrm{B}}$ is Boltzmann's constant, and $m_{^{13}\mathrm{CO}}$ is the mass of the $^{13}$CO molecule.
    
    The isothermal sound speed is
    \begin{equation}
    	c_s=\sqrt{\frac{k_{\mathrm{B}}T_{\mathrm{ex}}}{\mu m_{\mathrm{H}}}},
    \end{equation}
    
    \noindent where $\mu=2.33$ is the mean molecular weight per free particle for cold molecular gas of approximately solar composition \citep{2008A&A...487..993K}.
    
    The three-dimensional velocity dispersion is calculated as
    \begin{equation}
    	\sigma_{\mathrm{3D}}=\sqrt{3\left(c_s^2+\sigma_{\mathrm{NT}}^2\right)},
    \end{equation}
    and the $^{13}$CO Mach number is defined as
    \begin{equation}
    	\mathcal{M}_{^{13}\mathrm{CO}}=\frac{\sigma_{\mathrm{NT}}}{c_s}.
    \end{equation}
        
	The analysis yields a mean non-thermal velocity dispersion of 
	$\sigma_{\mathrm{NT}}=0.9\pm0.02$~\kms, where the quoted uncertainty is the standard error of the mean. The median and standard deviation are both 0.7~\kms. 
	These values are larger than those reported by \citet{2012ApJ...756...76W}, who found mean and median values of 0.5~\kms\ and a standard deviation of 0.3~\kms. 
	{From the definition of $\sigma_{\mathrm{NT}}$, this difference is expected to be driven mainly by the broader $^{13}$CO line widths measured in the present sample, because the characteristic $T_{\mathrm{ex}}$ values are very similar between the two studies and therefore introduce only a limited difference in the thermal correction term.}
	
	At the velocity-component level, the mean and median Mach numbers are 
	$\mathcal{M}_{^{13}\mathrm{CO}}=4.7$ and 3.7, respectively, again exceeding the mean value of 3.1 reported by \citet{2012ApJ...756...76W}. 
	{Since $\mathcal{M}_{^{13}\mathrm{CO}}=\sigma_{\mathrm{NT}}/c_s$ and the sound speed depends on $T_{\mathrm{ex}}$, the comparable excitation temperatures in the two samples imply that the higher Mach numbers found here likewise arise primarily from the larger non-thermal velocity dispersions rather than from a substantially different thermal scale. 
	As discussed above, the broader source-averaged line profiles may partly reflect the wider effective spatial sampling of the TRAO observations relative to central single-point measurements, in addition to possible differences in sample composition.} The mean three-dimensional velocity dispersion is $\sigma_{\mathrm{3D}}=1.5\pm0.02$~\kms. The relevant distributions are shown in Figure~\ref{fig:width}, and the complete parameter listings are provided in Tables~\ref{tab:2} and \ref{tab:3} in Appendix~A.

    Overall, the expanded sample confirms that supersonic non-thermal motions are widespread in PGCCs. The $\tau_{^{13}\mathrm{CO}}$ distribution is approximately lognormal, whereas the $N_{\mathrm{H_2-CO}}$ distribution is not. The latter result should be interpreted with caution, because $N_{\mathrm{H_2-CO}}$ depends directly on the assumption of a constant $[\mathrm{H}_2]/[^{13}\mathrm{CO}]$ ratio.
      
 	\section{Discussion}\label{sec:Discussion} 
 	
 	\subsection{Methodological considerations for source-averaged gas--dust comparisons}
 	
 	\begin{figure}[htbp]
 		\centering
 		\includegraphics[width=\linewidth]{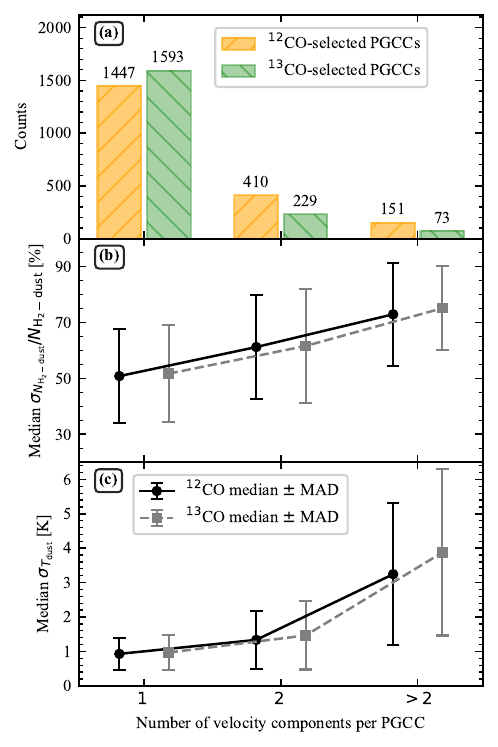}
 		\caption{{Catalog dust-parameter uncertainties for different CO velocity-component multiplicity bins. Panel (a) gives the numbers of PGCCs with one, two, and more than two detected $^{12}$CO or $^{13}$CO components. Panels (b) and (c) show the median relative uncertainty $\sigma_{N_{\mathrm{H_2-dust}}}/N_{\mathrm{H_2-dust}}$ and the median dust-temperature uncertainty $\sigma_{T_{\mathrm{dust}}}$ in the same multiplicity bins. Black circles and gray squares represent the $^{12}$CO- and $^{13}$CO-selected samples. Error bars denote median absolute deviations.}}
 		\label{fig:Lim2}
 	\end{figure}
 	
 	{In addition to the sparse discrete CO sampling discussed in Section~\ref{sec:observation}, source-averaged gas--dust comparisons may be affected by catalog dust uncertainties and by line-of-sight velocity complexity. 
 	Figure~\ref{fig:Lim2} examines whether these effects are statistically related by comparing the catalog dust-parameter uncertainties among PGCCs grouped according to the number of detected CO velocity components. 
 	For both the $^{12}$CO- and $^{13}$CO-selected samples, the median relative uncertainty in $N_{\mathrm{H_2-dust}}$ and the median uncertainty in $T_{\mathrm{dust}}$ increase from the single-component category toward higher component multiplicities, with the largest values occurring in the $>2$-component bin. The relatively large median absolute deviations indicate substantial source-to-source scatter.
 	Nevertheless, both the relative uncertainty in $N_{\mathrm{H_2-dust}}$ and the uncertainty in $T_{\mathrm{dust}}$ increase systematically with velocity-component multiplicity. This result indicates that dust properties are generally less well constrained toward kinematically more complex sightlines, although the underlying cause cannot be uniquely identified from the present analysis.}
 	
 	\begin{figure*}[htbp]
 		\centering
 		\includegraphics[width=\linewidth]{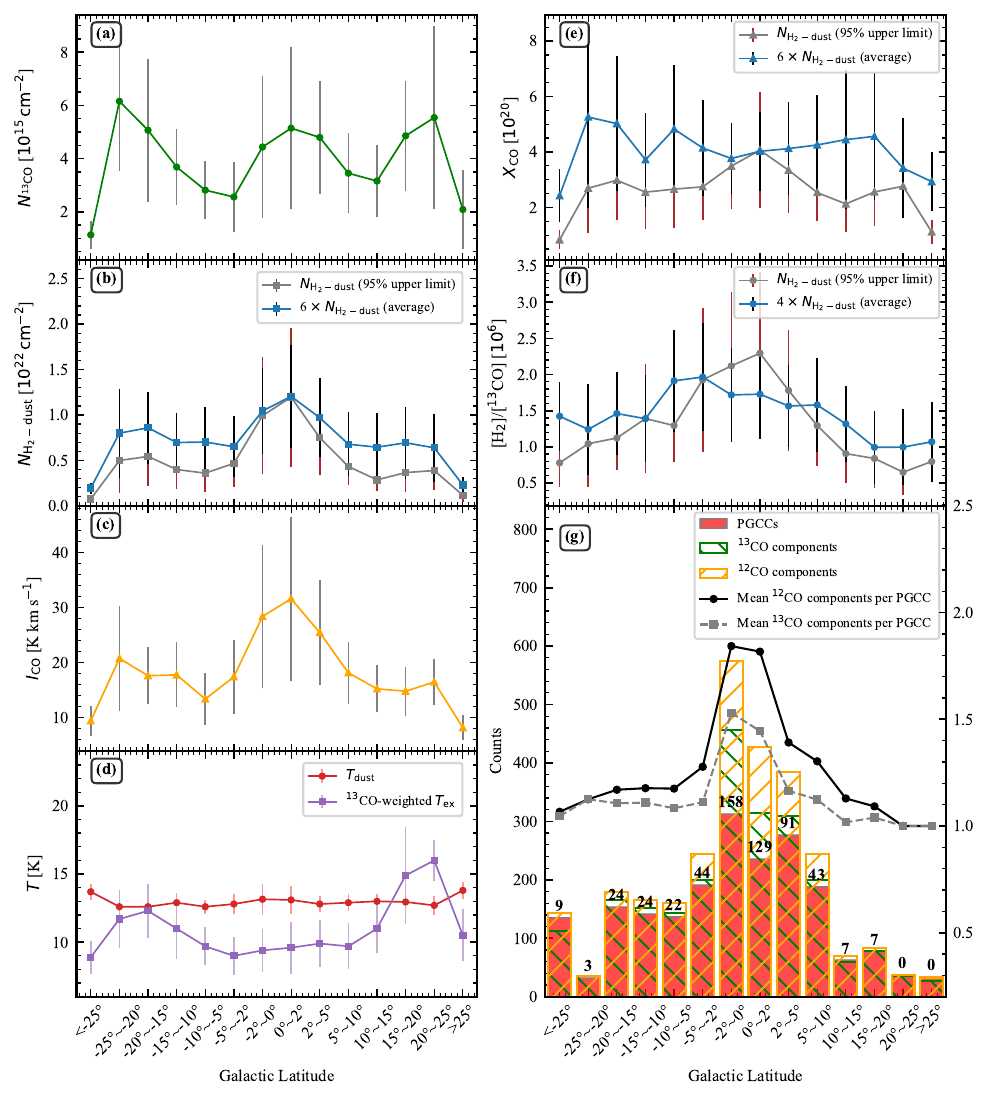} 		
 		\caption{{Galactic-latitude dependence of source-averaged gas--dust quantities and velocity-component statistics. Panels (a)--(f) show the binned median trends of (a) $N_{^{13}\mathrm{CO}}$, (b) $N_{\mathrm{H_2-dust}}$, (c) $I_{^{12}\mathrm{CO}}$, (d) the catalog dust temperature together with the $^{13}$CO-intensity-weighted excitation temperature, (e) $X_{\mathrm{CO}}$, and (f) $[\mathrm{H_2}]/[^{13}\mathrm{CO}]$. Error bars indicate the corresponding median absolute deviations. In panels (b), (e), and (f), gray symbols use the 95\% upper confidence limit of $N_{\mathrm{H_2-dust}}$, while blue symbols show the nominal catalog estimate multiplied by constant factors for display only, as indicated in the legends. Panel (g) presents the Galactic-latitude distributions of PGCCs, $^{13}$CO velocity components, and $^{12}$CO velocity components, together with the mean numbers of $^{12}$CO components per PGCC and $^{13}$CO components per $^{13}$CO-detected PGCC. Numbers above the bars denote the counts of PGCCs with multiple $^{12}$CO velocity components in each latitude bin.}}
 		\label{fig:Lim1}
 	\end{figure*}
 	
 	{The Galactic-latitude dependence of these effects is summarized in Figure~\ref{fig:Lim1}. 
 	Panel (g) shows that the mean number of detected velocity components per source increases toward lower Galactic latitudes for both isotopologues, with a stronger enhancement in $^{12}$CO. The numbers above the bars further demonstrate that PGCCs with multiple $^{12}$CO components are concentrated toward the Galactic plane. Consistent with the known concentration of molecular material at low Galactic latitude \citep{2001ApJ...547..792D,2015ARA&A..53..583H}, panels (b) and (c) show corresponding enhancements in $N_{\mathrm{H_2-dust}}$ and $I_{^{12}\mathrm{CO}}$ near the plane. These behaviors indicate that low-latitude sightlines are more susceptible to line-of-sight blending and component confusion than high-latitude sightlines.}
 	
 	{This complexity is particularly relevant for interpreting $X_{\rm CO}$, but two distinct effects should be separated. Radiative-transfer effects, including optical-depth saturation and self-absorption in $^{12}$CO, affect how the summed $I_{^{12}{\rm CO}}$ scales with the molecular gas column density. Line-of-sight blending, in contrast, affects whether the velocity-resolved CO components correspond to the same material represented by the catalog integrated dust SED quantities. In the source-averaged definition adopted here, the presence of additional $^{12}$CO components does not by itself imply a larger $X_{\rm CO}$, because their integrated intensities enter directly into the denominator. However, in kinematically crowded sightlines, the summed $I_{^{12}{\rm CO}}$ may not increase in proportion to the total gas column if part of the $^{12}$CO emission is affected by optical-depth effects, self-absorption, or saturation \citep{2013ARA&A..51..207B}.  At the same time, the dust-derived $N_{\rm H_2-dust}$ may include emission from multiple structures along the line of sight that are not recovered proportionally by the CO decomposition. The low-latitude enhancement of source-averaged $X_{\rm CO}$ should therefore be interpreted cautiously, as it may partly reflect radiative-transfer effects in $^{12}$CO and/or imperfect gas--dust association along complex sightlines, rather than a purely intrinsic variation in the conversion factor.}
 	
 	{A related caveat applies to the source-averaged [H$_2$]/[$^{13}$CO] ratio. In complex low latitude sightlines, the dust-derived H$_2$ column density can include contributions from structures that are not recovered proportionally in the summed $N_{^{13}{\rm CO}}$, especially when multiple components overlap or weak $^{13}$CO emission is difficult to separate robustly. Panels (e) and (f) of Figure~\ref{fig:Lim1} further show that adopting the 95\% upper confidence limit of $N_{\rm H_2-dust}$ strengthens the low-latitude behavior of both $X_{\rm CO}$ and [H$_2$]/[$^{13}$CO] relative to the nominal catalog estimate. This comparison indicates that the apparent Galactic-latitude dependence is sensitive not only to gas-side complexity, but also to the adopted dust-side catalog value. For this reason, we retain the nominal PGCC catalog $N_{\mathrm{H_2-dust}}$ values as the baseline choice in the main analysis.}
 	
 	{Taken together, the source-averaged gas--dust comparisons are subject to several distinct limitations.
 	Radiative-transfer effects, including optical-depth saturation and self-absorption, primarily influence the relation between CO intensity and gas column density. Line-of-sight blending mainly affects the correspondence between molecular-gas tracers and dust-derived quantities. These effects are further compounded by the discrete spatial sampling of the TRAO observations and by non-negligible uncertainties in the catalog dust parameters. The discussion below is therefore restricted to trend-level interpretations rather than strict quantitative calibration of $X_{\mathrm{CO}}$ or $[\mathrm{H_2}]/[^{13}\mathrm{CO}]$.}
 	
	\subsection{Gas--dust relations and possible CO depletion trends in PGCCs} \label{abundance}
	
	\begin{figure*}[htbp]
		\centering
		\includegraphics[width=\linewidth]{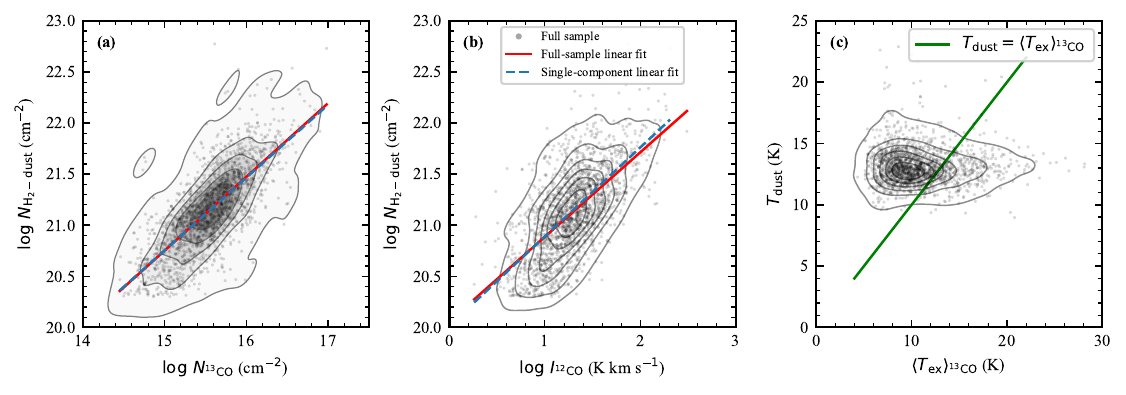}
		\caption{{Source-averaged gas--dust comparisons using the fixed-$\beta=2$ dust quantities from the PGCC catalog. Panels (a) and (b) show $N_{\mathrm{H_2-dust}}$ versus $N_{^{13}\mathrm{CO}}$ and $I_{^{12}\mathrm{CO}}$, respectively. Gray points and contours represent the source sample in each panel, and red solid lines show logarithmic linear fits to the full samples. Blue dashed lines, shown only in panels (a) and (b), give the corresponding single-component fits used to test the influence of multi-component sightlines, the subsets are defined by one detected $^{13}$CO component in panel (a) and one detected $^{12}$CO component in panel (b). Panel (c) compares the catalog dust temperature, $T_{\mathrm{dust}}$, with the $^{13}$CO-intensity-weighted excitation temperature, $\langle T_{\mathrm{ex}} \rangle_{^{13}\mathrm{CO}}$; the green line marks equality. }}
		
		\label{fig:gas_dust}
	\end{figure*}
	
	\begin{figure*}[htbp]
		\centering
		\includegraphics[width=0.9\linewidth]{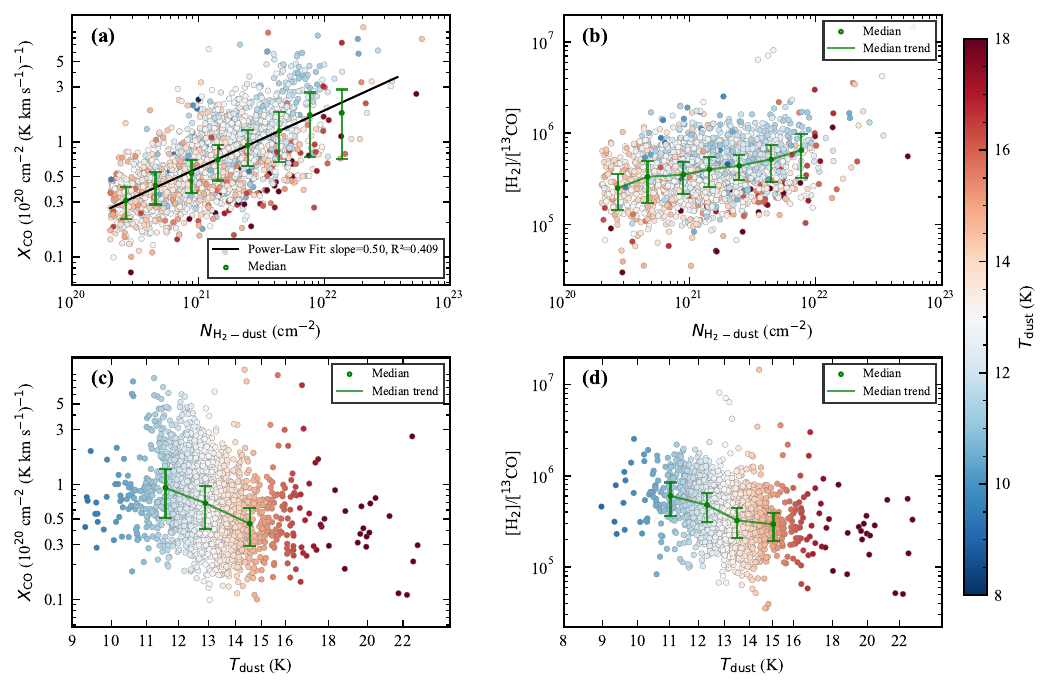}
		\caption{Variations of molecular gas parameters with dust-derived H$_2$ column density and dust temperature. Panel (a) shows $X_{\mathrm{CO}}$ versus $N_{\mathrm{H_2-dust}}$, panel (b) shows $[\mathrm{H_2}]/[^{13}\mathrm{CO}]$ versus $N_{\mathrm{H_2-dust}}$, panel (c) shows $X_{\mathrm{CO}}$ versus $T_{\mathrm{dust}}$, and panel (d) shows $[\mathrm{H_2}]/[^{13}\mathrm{CO}]$ versus $T_{\mathrm{dust}}$. The green symbols and connecting lines indicate the binned median trends.}
		\label{fig:trends}
	\end{figure*}
	
	Observational studies have shown that CO molecules can freeze out onto dust grains in cold, dense environments, thereby affecting the chemical and physical evolution of molecular clouds \citep{2002A&A...389L...6B,2002ApJ...570L.101B}.
	The PGCC sample, characterized by low dust temperatures and a broad range of dust-derived H$_2$ column densities, is therefore well suited for examining gas--dust relations that may be relevant to this process.
	Previous work by \citet{2013ApJ...775L...2L} further showed that the gaseous CO abundance, depletion factor, and CO-to-H$_2$ conversion factor in PGCCs correlate with several physical parameters, suggesting that CO abundance variations may provide useful empirical information on the physical evolution of cold clumps.	
	
	In this subsection, all gas--dust comparisons are performed at the source-averaged level. 
	For each PGCC, the dust-derived quantities $N_{\mathrm{H_2-dust}}$ and $T_{\mathrm{dust}}$ are taken directly from the PGCC catalog using the fixed $\beta=2$ values adopted in Section~\ref{sec:observation}. 
	The quantity $N_{^{13}\mathrm{CO}}$ is summed over all detected $^{13}$CO components, while $I_{^{12}\mathrm{CO}}$ is summed over all detected $^{12}$CO components, including sources without detected $^{13}$CO emission. 
	Comparisons involving $\langle T_{\mathrm{ex}} \rangle_{^{13}\mathrm{CO}}$ are restricted to sources with detected $^{13}$CO emission, for which the source-averaged excitation temperature is defined as the $^{13}$CO-intensity-weighted average over all identified components. 
	These comparisons should therefore be interpreted as source-averaged, clump-scale relations rather than strictly beam-matched measurements.
	
	{Panels (a) and (b) of Figure~\ref{fig:gas_dust} show statistically significant positive correlations between $N_{\mathrm{H_2-dust}}$ and both $N_{^{13}\mathrm{CO}}$ and $I_{^{12}\mathrm{CO}}$, although both relations remain moderate in strength. 
	For the full source sample, linear fits in logarithmic space, equivalent to power-law relations in the original variables, are given by
	\begin{align}
		\log N_{\mathrm{H_2-dust}} &= 0.83(\pm0.02)\,\log I_{^{12}\mathrm{CO}} + 20.06(\pm0.03), \\
		\log N_{\mathrm{H_2-dust}} &= 0.72(\pm0.01)\,\log N_{^{13}\mathrm{CO}} + 9.93(\pm0.23).
	\end{align}
	The corresponding coefficients of determination are $R^2=0.4$ for the $I_{^{12}\mathrm{CO}}$ relation and $R^2=0.6$ for the $N_{^{13}\mathrm{CO}}$ relation. 
	The fits obtained for the corresponding single-component subsets, shown by the dashed lines in Panels (a) and (b), closely follow the full-sample relations. 
	This agreement suggests that the main gas--dust correlations are not appreciably altered by the inclusion of multi-component sightlines.}
	
	{The $N_{^{13}\mathrm{CO}}$ relation is empirically tighter than the $I_{^{12}\mathrm{CO}}$ relation. 
	Although this difference is consistent with the generally lower optical depth of $^{13}$CO relative to $^{12}$CO, the present source-averaged analysis does not uniquely establish its physical origin. 
	We therefore regard it as a tighter observational correlation with $N_{\mathrm{H_2-dust}}$, rather than as evidence that $^{13}$CO is intrinsically a more faithful H$_2$ tracer in this sample.}
	
	{Panel (c) of Figure~\ref{fig:gas_dust} compares the catalog dust temperature, $T_{\mathrm{dust}}$, with the source-averaged $^{13}$CO-intensity-weighted excitation temperature, $\langle T_{\mathrm{ex}} \rangle_{^{13}\mathrm{CO}}$, for the $^{13}$CO-detected sources. The latter has a mean of 10.8~K and a median of 10.0~K, both below the typical catalog dust temperature of $\sim$13--14~K. About 80\% of the sources have $T_{\mathrm{dust}}>\langle T_{\mathrm{ex}} \rangle_{^{13}\mathrm{CO}}$, broadly consistent with the trend reported by \citet{2012ApJ...756...76W}. This offset should be interpreted cautiously because the two temperatures are not measured in the same way: $T_{\mathrm{dust}}$ is derived from dust SED fitting, whereas $\langle T_{\mathrm{ex}} \rangle_{^{13}\mathrm{CO}}$ is weighted by the $^{13}$CO emission from the detected velocity components. The observed difference may therefore reflect a combination of tracer-dependent weighting, unresolved temperature structure, beam and sampling differences, and possible gas--dust thermal decoupling. Thus, the present source-averaged comparison does not by itself identify a unique physical origin for the temperature offset.}
	
	{Figure~\ref{fig:trends} extends the gas--dust comparison to the two CO abundance-sensitive quantities, $X_{\mathrm{CO}}$ and $[\mathrm{H_2}]/[^{13}\mathrm{CO}]$. 
	Panel (a) shows that $X_{\mathrm{CO}}$ increases with $N_{\mathrm{H_2-dust}}$, which is the clearest trend among the relations considered here. 
	A similar but weaker increase is present for $[\mathrm{H_2}]/[^{13}\mathrm{CO}]$ in Panel (b). Panels (c) and (d) further show that both quantities tend to be higher toward lower $T_{\mathrm{dust}}$, although these temperature-dependent trends are moderate and dispersed. Given the source-averaged nature of the analysis and the limitations discussed above, the observed variations in $X_{\mathrm{CO}}$ and $[\mathrm{H_2}]/[^{13}\mathrm{CO}]$ should be regarded as empirical trends compatible with enhanced CO depletion or freeze-out, rather than as a direct demonstration of that process.}
	
	{Overall, the source-averaged gas--dust comparisons reveal moderate but statistically significant correlations. The single-component subsets yield fits that are close to those of the full samples, indicating that multi-component sightlines do not substantially drive the main relations. The observed trends therefore appear to be representative of the broader PGCC sample rather than being dominated by kinematically complex sources. Nevertheless, their physical interpretation remains non-unique. The tighter $N_{^{13}\mathrm{CO}}$--$N_{\mathrm{H_2-dust}}$ relation, the dust--gas temperature offset, and the variations in $X_{\mathrm{CO}}$ and $[\mathrm{H_2}]/[^{13}\mathrm{CO}]$ may involve tracer-dependent weighting, line-of-sight effects, radiative-transfer limitations, and abundance variations. Thus, although the abundance-sensitive trends are compatible with enhanced CO depletion in colder and higher-column-density PGCCs, they should be interpreted as empirical source-averaged trends rather than as evidence for a unique physical mechanism.}
	  
    \subsection{Non-thermal motions}
    
    \begin{figure}[htbp]
    	\centering
    	\includegraphics[width=0.9\linewidth]{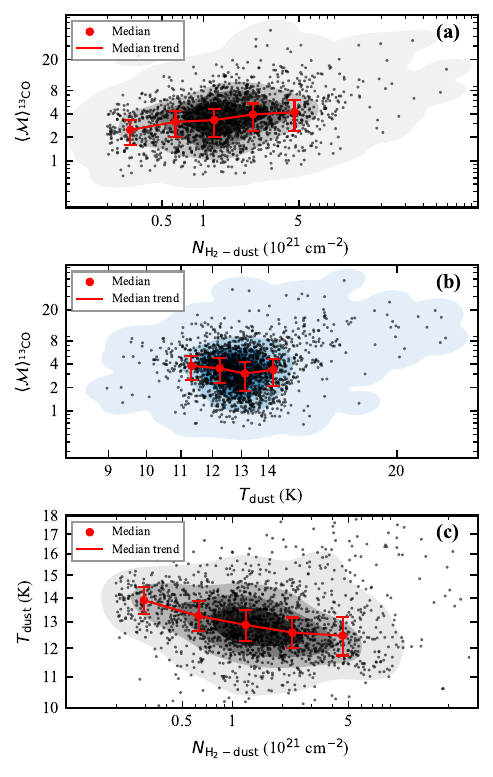}    	
    	\caption{The source-averaged $^{13}$CO intensity-weighted Mach number, $\langle\mathcal{M}\rangle_{^{13}\mathrm{CO}}$, versus the dust-derived H$_2$ column density, $N_{\mathrm{H_2-dust}}$ (a), and the catalog dust temperature, $T_{\mathrm{dust}}$ (b). Panel (c) shows $T_{\mathrm{dust}}$ versus $N_{\mathrm{H_2-dust}}$ for the same sample. Black points represent individual PGCCs, red symbols indicate median values in logarithmic bins, and error bars show the corresponding median absolute deviations. Gray and blue contours trace the kernel-density distribution.}
    	\label{fig:nt}
    \end{figure}  
    
    {Non-thermal motions are a key component of the dynamical state of PGCCs. Here we consider the source-averaged $^{13}$CO-intensity-weighted Mach number, $\langle \mathcal{M} \rangle_{^{13}\mathrm{CO}}$, for the 1895 PGCCs with detected $^{13}$CO emission. The mean and median values are 4.3 and 3.6, respectively, indicating that supersonic non-thermal motions are widespread in the sample. We compare $\langle \mathcal{M} \rangle_{^{13}\mathrm{CO}}$ with the dust-derived H$_{2}$ column density, $N_{\mathrm{H_2-dust}}$, and the catalog dust temperature, $T_{\mathrm{dust}}$, as source-level descriptors of the clump environment. Here $T_{\mathrm{dust}}$ is used only as an independent catalog-scale dust property; the gas temperature entering the Mach-number calculation is already accounted for through the sound speed derived from $T_{\mathrm{ex}}$.}
    
    {Figure~\ref{fig:nt} summarizes these source-level trends. Panel (a) shows a mild increase in the binned median $\langle \mathcal{M} \rangle_{^{13}\mathrm{CO}}$ with increasing $N_{\mathrm{H_2-dust}}$, although the individual sources show substantial scatter. Panel (b) shows no comparably clear monotonic dependence on $T_{\mathrm{dust}}$. Panel (c) confirms that, within the same $^{13}$CO-detected sample, higher-column-density sources tend to have lower catalog dust temperatures. Thus, the source-averaged Mach number shows a weak dependence on column density but is not simply controlled by the dust temperature.}
    
    {The main result is that PGCCs exhibit widespread supersonic non-thermal motions, with a mild enhancement of $\langle \mathcal{M} \rangle_{^{13}\mathrm{CO}}$ toward higher dust-derived column density. The physical origin of this behavior is not uniquely determined from the present source-averaged data, because internal gas dynamics, environmental turbulence, and line-of-sight complexity may all contribute. Spatially resolved observations of selected PGCCs, preferably with complementary dense-gas tracers, will be needed to identify the dominant driver of the observed non-thermal motions.}

\section{Summary}\label{sec:summary}
	
	We present a systematic CO survey of 2008 Planck Galactic Cold Clumps (PGCCs), based on multi-beam single-pointing observations of the $^{12}$CO and $^{13}$CO ($J=1\!-\!0$) transitions obtained with the TRAO 14~m telescope. The resulting catalog includes both observed parameters, such as $V_{\mathrm{LSR}}$, FWHM, and $T_{\mathrm{mb}}$, and derived quantities, such as $T_{\mathrm{ex}}$, $\tau_{^{13}\mathrm{CO}}$, $N_{^{13}\mathrm{CO}}$, $\sigma_{\mathrm{NT}}$, and $\mathcal{M_\mathrm{^{13}CO}}$, providing a statistical description of the kinematic, thermal, and dynamical properties of PGCCs. Compared with the 674-source CO survey of \citet{2012ApJ...756...76W}, the present 2008-source sample provides an improved statistical basis for investigating the physical conditions of cold Galactic clumps. The principal results are summarized as follows.
	
	\begin{enumerate}
		\item The single-pointing survey, conducted between December 2015 and March 2018, detected $^{12}$CO emission toward all 2008 PGCCs, yielding 2784 individual velocity components, while $^{13}$CO emission was detected toward 1895 PGCCs, corresponding to 2291 velocity components. Multi-component spectral features occur mainly at low Galactic latitudes ($|b| < 5^\circ$). The close correspondence between the $^{12}$CO and $^{13}$CO line-center velocities indicates that the two tracers generally probe the same kinematic structures.
		
		\item Among the velocity components with both $^{12}$CO and $^{13}$CO detections, the corresponding source-averaged $^{12}$CO peak main-beam brightness temperature, obtained from Gaussian fitting to the multi-beam averaged spectrum, has a mean of 7.1~K and a median of 6.3~K. The derived excitation temperature is close to 10~K on average, comparable to the mean value of 10.1~K found by \citet{2012ApJ...756...76W}. The $^{13}$CO optical depth has a mean of 0.5 and a median of 0.4, both lower than the mean value of 0.9 reported in the same study. In addition, the $^{13}$CO optical-depth distribution is approximately lognormal.
		
		\item The median FWHM values are 2.5~\kms\ for $^{12}$CO and 1.6~\kms\ for $^{13}$CO emission components, with corresponding mean values of 3.0 and 2.0~\kms. These line widths are broader than those reported by \citet{2012ApJ...756...76W}. Adopting the commonly used criterion $\mathrm{FWHM}(^{13}\mathrm{CO}) > 1.3$~\kms \citep{1983ApJ...266..309M}, 63\% of the $^{13}$CO components in our sample would be classified as high-mass clump candidates, compared with 36\% in the sample of \citet{2012ApJ...756...76W}. This result suggests that the PGCCs in the present sample tend to show stronger non-thermal motions than those in the earlier, smaller sample.
				
		\item Comparison with the dust-derived properties shows moderate gas--dust correlations. The relation between $N_{^{13}{\rm CO}}$ and the dust-derived H$_2$ column density is somewhat tighter than that involving $I_{^{12}{\rm CO}}$, but this should be interpreted as an empirical source-averaged correlation rather than as proof that $^{13}$CO is intrinsically a uniquely direct H$_2$ tracer. The source-averaged quantities $X_{\rm CO}$ and [H$_2$]/[$^{13}$CO] are likewise derived from the adopted dust-based H$_2$ column density and should therefore be regarded as empirical quantities whose absolute values and distribution properties may vary with the choice of catalog dust estimate.
		
		\item {The abundance-sensitive quantities show trends that are compatible with enhanced CO depletion or freeze-out in colder and higher-column-density PGCCs. In the enlarged source-averaged sample, $X_{\rm CO}$ shows a clearer increase with dust-derived H$_2$ column density, whereas the corresponding trend in [H$_2$]/[$^{13}$CO] is weaker and more scattered. Both quantities also show higher values toward the low-$T_{\rm dust}$ end. These relations remain moderate in strength and are best interpreted as empirical trends, not as evidence for a distinct threshold or a unique physical mechanism.}
			
		\item The enlarged sample confirms that supersonic non-thermal motions remain widespread in PGCCs. For the source-averaged $^{13}$CO-intensity-weighted Mach number, the mean and median values are 4.3 and 3.6, respectively. A mild increase is seen with dust-derived H$_2$ column density, whereas no comparably clear monotonic trend is found with dust temperature. The physical origin of this behavior remains uncertain in the present source-averaged analysis. Future higher-resolution mapping of carefully selected, preferably single-velocity sources, together with dense-gas tracers such as NH$_3$, will be important 
		for identifying the dynamical processes operating within individual PGCCs.
	\end{enumerate}    
       
    \begin{acknowledgments}
    	This work was primarily supported by the National Key R\&D Program of China (Grant Nos. 2022YFA1603103, 2022YFA1603101, 2023YFA1608002, and 2022YFA1603100). Additional support was received from the Tianshan Talent Training Program (2024TSYCTD0013) and the National Natural Science Foundation of China (Grants 12173075, 12373029, 12403033, 12073061, and 12122307). Further funding was provided by the CAS ``Light of West China" Program (XBZG-ZDSYS-202212), the Xinjiang Key Laboratory of Radio Astrophysics (2023D04033), the Tianshan Talent Program of Xinjiang Uygur Autonomous Region (2022TSY-CLJ0005), and the Central Guidance Fund for Local Science and Technology Development (ZYYD2025ZYZ3). G.W. acknowledges support from the Xinjiang Tianchi Talents Program and the Youth Innovation Promotion Association of CAS. Based on observations made with the TRAO 14 m telescope, we sincerely thank the TRAO staff for their long-term support and assistance. This work is dedicated to the memory of Professor Yuefang Wu, a pioneer who employed spectral lines of CO molecules to study PGCCs﻿.
    \end{acknowledgments}
    \clearpage
     
    \appendix
    In this appendix, we provide the full tabulated data and statistical summaries for the sources observed in this survey. Table \ref{tab:line_parameters} lists the direct Gaussian fitting results of the $^{12}$CO and $^{13}$CO emission lines, including the integrated intensities ($I_{\rm CO}$), central LSR velocities (\(V_{\rm LSR}\)), FWHM linewidths, main-beam brightness temperatures ($T_{\rm mb}$), and their corresponding formal $1\sigma$ fitting uncertainties. Table \ref{tab:2} summarizes the derived physical parameters for each individual source, including the excitation temperature ($T_{\rm ex}$), $^{13}$CO optical depth ($\tau_\mathrm{^{13}CO}$), $^{13}$CO and H$_{2}$ column densities, non-thermal velocity dispersion ($\sigma_{\rm NT}$), isothermal sound speed ($c_{\rm s}$), 3D velocity dispersion ($\sigma_{\rm 3D}$), $^{12}$CO/$^{13}$CO integrated intensity ratio, and the Mach number ($\mathcal{M}_{\rm ^{13}CO}$). Table \ref{tab:3} presents the statistical analysis of these derived parameters for the surveyed Planck Galactic Cold Clumps (PGCCs). For each parameter, we compute statistical metrics including the sample size ($N$), minimum, maximum, mean, median, standard deviation (Std dev), and median absolute deviation (MAD). Furthermore, we perform lognormal fitting to characterize the parameter distributions. The rows labeled $\mu$ and $\sigma$ represent the mean and standard deviation of the natural logarithm of the fitted distributions, respectively, while the $P$ value denotes the probability from the goodness-of-fit test. As noted in the table footnote, only parameters involving the $^{13}$CO component were used for the statistical analysis, and the MAD represents the median absolute deviation from the median.\\

    \onecolumngrid  \section{Information of \ce{^{12}CO} and \ce{^{13}CO} emission from PGCCs} \label{}
    
	\begin{table*}[htbp]
		\centering
		\small 
		\setlength{\tabcolsep}{2pt}
		\begin{threeparttable}
			\caption{Line parameters of \ce{^{12}CO} and \ce{^{13}CO} emission from PGCCs}
			\label{tab:line_parameters}
			\begin{tabular}{l *{8}{c}}
				\toprule
				Name & 
				\multicolumn{1}{c}{$I_{^{12}\text{CO}}$} & 
				\multicolumn{1}{c}{$V_{\mathrm{lsr}}(^{12}\mathrm{CO})$} & 
				\multicolumn{1}{c}{FWHM($^{12}$CO)} & 
				\multicolumn{1}{c}{$T_{\textrm{mb}}(^{12}\textrm{CO})$} & 
				\multicolumn{1}{c}{$I_{^{13}\text{CO}}$} & 
				\multicolumn{1}{c}{$V_{\mathrm{lsr}}(^{13}\mathrm{CO})$} & 
				\multicolumn{1}{c}{FWHM($^{13}$CO)} & 
				\multicolumn{1}{c}{$T_{\textrm{mb}}(^{13}\textrm{CO})$} \\
				\multicolumn{1}{c}{} & 
				\multicolumn{1}{c}{(K \kms)} & 
				\multicolumn{1}{c}{(\kms)} & 
				\multicolumn{1}{c}{(\kms)} & 
				\multicolumn{1}{c}{(K)} & 
				\multicolumn{1}{c}{(K \kms)} & 
				\multicolumn{1}{c}{(\kms)} & 
				\multicolumn{1}{c}{(\kms)} & 
				\multicolumn{1}{c}{(K)} \\
				\midrule
				G000.01+18.02 & 11.71 (0.32) & 5.83 (0.01) & 0.87 (0.03) & 12.68 (1.12) & 3.38 (0.10) & 5.85 (0.01) & 0.59 (0.02) & 5.37 (0.21) \\
				G000.06-00.69 & 87.03 (1.10) & 15.38 (0.05) & 8.04 (0.11) & 10.16 (1.33) & 27.30 (0.30) & 15.70 (0.03) & 6.17 (0.08) & 4.16 (0.19) \\
				G000.08+00.20 & 28.13 (2.93) & -5.17 (0.40) & 6.83 (0.67) & 3.87 (0.68) & 7.76 (0.34) & -5.30 (0.10) & 4.37 (0.21) & 1.67 (0.27) \\
				G000.08+00.20 & 28.49 (2.93) & 1.44 (0.19) & 4.54 (0.38) & 5.90 (0.68) & 8.76 (0.29) & 0.83 (0.05) & 2.96 (0.11) & 2.78 (0.27) \\
				G000.08+00.20 & 27.93 (1.82) & -34.37 (0.31) & 9.44 (0.75) & 2.78 (0.68) &  &  &  &  \\
				G000.12+21.77 & 18.09 (0.12) & 0.61 (0.00) & 1.30 (0.01) & 13.05 (0.36) & 3.99 (0.03) & 0.29 (0.00) & 0.50 (0.00) & 7.57 (0.14) \\
				G000.18-00.50 & 36.75 (0.95) & -3.41 (0.07) & 5.64 (0.17) & 6.13 (1.33) & 18.40 (0.26) & -2.77 (0.03) & 4.58 (0.07) & 3.76 (0.19) \\
				G000.18-00.50 & 61.74 (0.93) & 17.29 (0.04) & 5.15 (0.09) & 11.25 (1.33) & 21.89 (0.23) & 17.24 (0.02) & 3.61 (0.04) & 5.70 (0.19) \\
				G000.21-00.97 & 71.55 (1.12) & 15.11 (0.05) & 6.93 (0.14) & 9.70 (1.28) & 15.41 (0.27) & 15.37 (0.03) & 3.87 (0.09) & 3.74 (0.21) \\
				G000.24+11.71 & 10.57 (0.31) & 2.65 (0.01) & 0.77 (0.03) & 12.88 (0.59) & 3.68 (0.09) & 2.66 (0.01) & 0.53 (0.01) & 6.57 (0.21) \\
				G000.28+11.39 & 14.25 (0.33) & 3.45 (0.02) & 1.64 (0.05) & 8.15 (0.82) & 3.09 (0.06) & 3.29 (0.01) & 0.52 (0.01) & 5.52 (0.14) \\
				G000.44+10.29 & 22.67 (0.44) & 4.58 (0.01) & 1.35 (0.03) & 15.76 (1.27) & 6.00 (0.11) & 4.60 (0.01) & 0.92 (0.02) & 6.11 (0.19) \\
				G000.49+11.38 & 22.11 (0.43) & 2.98 (0.01) & 1.52 (0.03) & 13.66 (1.18) & 7.25 (0.11) & 3.17 (0.01) & 0.94 (0.02) & 7.24 (0.19) \\
				G000.65+22.32 & 22.06 (0.25) & 0.62 (0.02) & 2.83 (0.03) & 7.32 (0.27) & 1.34 (0.06) & 1.66 (0.02) & 0.73 (0.04) & 1.72 (0.11) \\
				G000.66-00.82 & 56.73 (1.02) & 15.81 (0.07) & 7.89 (0.15) & 6.75 (1.24) & 11.89 (0.26) & 16.43 (0.08) & 7.19 (0.18) & 1.55 (0.12) \\
				G000.81+25.65 & 4.70 (0.26) & 0.51 (0.03) & 1.09 (0.07) & 4.05 (0.42) &  &  &  &  \\
				G000.93+07.12 & 20.33 (0.39) & 5.16 (0.02) & 2.54 (0.06) & 7.53 (0.83) & 3.83 (0.11) & 5.27 (0.03) & 2.16 (0.07) & 1.66 (0.13) \\
				G000.95+07.65 & 16.91 (0.72) & 5.96 (0.05) & 2.22 (0.11) & 7.17 (1.07) & 4.25 (0.11) & 5.79 (0.01) & 1.06 (0.03) & 3.78 (0.17) \\
				G001.03+15.88 & 17.69 (0.40) & 4.85 (0.01) & 1.16 (0.03) & 14.29 (1.25) & 4.32 (0.12) & 4.75 (0.01) & 0.71 (0.02) & 5.70 (0.24) \\
				G001.24+09.89 & 22.92 (0.45) & 4.03 (0.02) & 1.68 (0.04) & 12.80 (1.19) & 7.05 (0.13) & 3.96 (0.01) & 1.06 (0.02) & 6.24 (0.20) \\
				G001.30+06.61 & 14.87 (0.39) & 4.35 (0.02) & 1.40 (0.04) & 9.98 (0.56) & 3.13 (0.10) & 4.34 (0.02) & 0.97 (0.03) & 3.02 (0.18) \\
				G001.40+20.93 & 16.58 (0.15) & 0.60 (0.01) & 1.12 (0.01) & 13.90 (0.44) & 7.05 (0.04) & 0.62 (0.00) & 0.73 (0.00) & 9.02 (0.16) \\
				G001.46+06.94 & 7.81 (0.94) & 5.01 (0.13) & 2.11 (0.31) & 3.49 (1.05) & 1.18 (0.12) & 4.64 (0.09) & 1.54 (0.18) & 0.72 (0.12) \\
				G001.52+07.09 & 3.20 (0.52) & 3.52 (0.06) & 0.75 (0.14) & 4.00 (1.01) & 1.14 (0.08) & 3.42 (0.02) & 0.48 (0.04) & 2.22 (0.18) \\
				G001.62-00.01 & 121.62 (1.50) & 53.94 (0.10) & 16.73 (0.22) & 6.82 (1.27) & 16.58 (0.56) & 51.93 (0.23) & 13.75 (0.52) & 1.13 (0.18) \\
				G001.70+17.43 & 10.96 (0.32) & 4.90 (0.01) & 0.89 (0.03) & 11.59 (1.10) & 1.79 (0.10) & 4.97 (0.02) & 0.80 (0.05) & 2.10 (0.13) \\
				G001.72-00.40 & 32.85 (1.05) & 15.80 (0.10) & 7.52 (0.28) & 4.11 (1.23) & 6.69 (0.38) & 15.57 (0.16) & 5.73 (0.41) & 1.10 (0.18) \\
				G001.72-00.40 & 42.74 (0.81) & -31.64 (0.15) & 10.66 (0.30) & 3.76 (1.23) &  &  &  &  \\
				G001.77+06.96 & 11.61 (0.48) & 4.88 (0.02) & 0.77 (0.04) & 14.15 (0.92) & 4.56 (0.09) & 4.90 (0.01) & 0.69 (0.02) & 6.22 (0.19) \\
				G001.84+16.59 & 15.93 (0.41) & 5.94 (0.02) & 1.32 (0.04) & 11.29 (1.16) & 6.08 (0.12) & 5.82 (0.01) & 0.75 (0.02) & 7.63 (0.22) \\
				G002.12+17.45 & 14.71 (0.37) & 5.57 (0.01) & 1.20 (0.04) & 11.49 (1.07) & 3.78 (0.11) & 5.56 (0.01) & 0.76 (0.03) & 4.70 (0.22) \\
				G002.15+21.84 & 13.24 (0.16) & -0.40 (0.01) & 1.30 (0.02) & 9.59 (0.45) & 3.91 (0.06) & -0.40 (0.00) & 0.77 (0.01) & 4.80 (0.19) \\
				G002.21+17.63 & 19.27 (0.34) & 5.86 (0.01) & 1.18 (0.02) & 15.29 (1.08) & 4.38 (0.11) & 5.76 (0.01) & 0.76 (0.02) & 5.43 (0.21) \\
				G002.49+09.75 & 22.53 (0.52) & 2.95 (0.02) & 1.62 (0.04) & 13.10 (1.33) & 5.26 (0.15) & 2.89 (0.01) & 0.97 (0.03) & 5.09 (0.24) \\			
				\bottomrule
			\end{tabular}
			\begin{tablenotes}
				\item [†]This table is available in its entirety in machine-readable form in the online journal. A portion is shown here for guidance regarding its form and content. All observed parameters have been corrected for the main beam efficiency of the telescope.
			\end{tablenotes}
		\end{threeparttable}
	\end{table*}      
	\begin{table*}[htbp]
		\centering
		\small \setlength{\tabcolsep}{2pt}
		\begin{threeparttable}
			\caption{Derived parameters of surveyed PGCCs}
			\label{tab:2}
			\begin{tabular}{l *{12}{c}}
				\toprule
				Name &
				\multicolumn{1}{c}{Glon} & 
				\multicolumn{1}{c}{Glat} &
				\multicolumn{1}{c}{$V_\mathrm{lsr}$} & 
				\multicolumn{1}{c}{$T_\mathrm{ex}$} & 
				\multicolumn{1}{c}{$\tau_\mathrm{^{13}CO}$} & 
				\multicolumn{1}{c}{$N_\mathrm{^{13}CO}$} & 
				\multicolumn{1}{c}{$N_\mathrm{H_{2}}$} & 
				\multicolumn{1}{c}{$I_{^{12}\text{CO}}$/$I_{^{13}\text{CO}}$} & 
				\multicolumn{1}{c}{$\sigma_\mathrm{NT}$} & 
				\multicolumn{1}{c}{$c_{s}$}&
				\multicolumn{1}{c}{$\sigma_\mathrm{3D}$}&
				\multicolumn{1}{c}{$\mathcal{M}_{\rm ^{13}CO}$} \\
				\multicolumn{1}{c}{} &
				\multicolumn{1}{c}{(\textdegree)}&
				\multicolumn{1}{c}{(\textdegree)}&
				\multicolumn{1}{c}{(\kms)} &
				\multicolumn{1}{c}{(K)} & 
				\multicolumn{1}{c}{} &  
				\multicolumn{1}{c}{(10$^{15}$ cm$^{-2}$)} & 
				\multicolumn{1}{c}{(10$^{21}$ cm$^{-2}$)} & 
				\multicolumn{1}{c}{} & 
				\multicolumn{1}{c}{(\kms)} & 
				\multicolumn{1}{c}{(\kms)} & 
				\multicolumn{1}{c}{(\kms)} &
				\multicolumn{1}{c}{} \\
				\midrule
				G000.01+18.02 & 0.013 & 18.021 & 5.85 & 16.1 & 0.6 & 4.0 & 3.6 & 3.5 & 0.24 & 0.24 & 0.59 & 1.0 \\
				G000.06-00.69 & 0.063 & -0.689 & 15.70 & 13.6 & 0.5 & 28.0 & 24.9 & 3.2 & 2.62 & 0.22 & 4.56 & 11.9 \\
				G000.08+00.20 & 0.083 & 0.204 & -5.30 & 7.1 & 0.6 & 5.3 & 4.7 & 3.6 & 1.86 & 0.16 & 3.23 & 11.7 \\
				G000.08+00.20 & 0.083 & 0.204 & 0.83 & 9.2 & 0.6 & 7.2 & 6.4 & 3.3 & 1.25 & 0.18 & 2.19 & 6.9 \\
				G000.12+21.77 & 0.118 & 21.774 & 0.29 & 16.5 & 0.9 & 5.5 & 4.9 & 4.5 & 0.20 & 0.24 & 0.54 & 0.8 \\
				G000.18-00.50 & 0.183 & -0.499 & -2.77 & 9.4 & 1.0 & 17.6 & 15.7 & 2.0 & 1.95 & 0.18 & 3.38 & 10.6 \\
				G000.18-00.50 & 0.183 & -0.499 & 17.24 & 14.7 & 0.7 & 25.8 & 23.0 & 2.8 & 1.53 & 0.23 & 2.68 & 6.7 \\
				G000.21-00.97 & 0.209 & -0.968 & 15.37 & 13.1 & 0.5 & 15.1 & 13.5 & 4.6 & 1.64 & 0.22 & 2.87 & 7.6 \\
				G000.24+11.71 & 0.238 & 11.708 & 2.66 & 16.3 & 0.7 & 4.7 & 4.2 & 2.9 & 0.21 & 0.24 & 0.56 & 0.9 \\
				G000.28+11.39 & 0.284 & 11.385 & 3.29 & 11.5 & 1.1 & 3.6 & 3.2 & 4.6 & 0.22 & 0.20 & 0.51 & 1.1 \\
				G000.44+10.29 & 0.438 & 10.291 & 4.60 & 19.2 & 0.5 & 8.0 & 7.1 & 3.8 & 0.39 & 0.26 & 0.81 & 1.5 \\
				G000.49+11.38 & 0.487 & 11.384 & 3.18 & 17.1 & 0.8 & 9.9 & 8.8 & 3.0 & 0.39 & 0.25 & 0.80 & 1.6 \\
				G000.65+22.32 & 0.648 & 22.322 & 1.66 & 10.7 & 0.3 & 1.0 & 0.9 & 16.5 & 0.31 & 0.20 & 0.63 & 1.6 \\
				G000.66-00.82 & 0.663 & -0.822 & 16.43 & 10.1 & 0.3 & 8.7 & 7.8 & 4.8 & 3.05 & 0.19 & 5.30 & 16.1 \\
				G000.93+07.12 & 0.932 & 7.118 & 5.27 & 10.9 & 0.2 & 2.9 & 2.6 & 5.3 & 0.92 & 0.20 & 1.62 & 4.6 \\
				G000.95+07.65 & 0.945 & 7.652 & 5.79 & 10.5 & 0.7 & 4.0 & 3.6 & 4.0 & 0.45 & 0.19 & 0.84 & 2.3 \\
				G001.03+15.88 & 1.028 & 15.877 & 4.74 & 17.7 & 0.5 & 5.4 & 4.8 & 4.1 & 0.29 & 0.25 & 0.67 & 1.2 \\
				G001.24+09.89 & 1.240 & 9.894 & 3.96 & 16.2 & 0.7 & 8.9 & 7.9 & 3.3 & 0.44 & 0.24 & 0.88 & 1.8 \\
				G001.30+06.61 & 1.298 & 6.614 & 4.34 & 13.4 & 0.4 & 2.9 & 2.6 & 4.8 & 0.41 & 0.22 & 0.80 & 1.9 \\
				G001.40+20.93 & 1.396 & 20.933 & 0.62 & 17.3 & 1.0 & 11.0 & 9.8 & 2.4 & 0.30 & 0.25 & 0.68 & 1.2 \\
				G001.46+06.94 & 1.456 & 6.944 & 4.64 & 6.7 & 0.2 & 0.7 & 0.6 & 6.6 & 0.65 & 0.15 & 1.16 & 4.2 \\
				G001.52+07.09 & 1.522 & 7.085 & 3.42 & 7.2 & 0.8 & 0.9 & 0.8 & 2.8 & 0.20 & 0.16 & 0.45 & 1.2 \\
				G001.62-00.01 & 1.624 & -0.011 & 51.93 & 10.2 & 0.2 & 11.7 & 10.5 & 7.3 & 5.84 & 0.19 & 10.12 & 30.7 \\
				G001.70+17.43 & 1.695 & 17.430 & 4.97 & 15.0 & 0.2 & 1.7 & 1.5 & 6.1 & 0.33 & 0.23 & 0.70 & 1.4 \\
				G001.72-00.40 & 1.717 & -0.404 & 15.57 & 7.3 & 0.3 & 4.1 & 3.7 & 4.9 & 2.43 & 0.16 & 4.22 & 15.0 \\
				G001.77+06.96 & 1.772 & 6.959 & 4.90 & 17.6 & 0.6 & 5.9 & 5.2 & 2.5 & 0.28 & 0.25 & 0.66 & 1.1 \\
				G001.84+16.59 & 1.843 & 16.589 & 5.82 & 14.7 & 1.1 & 8.6 & 7.7 & 2.6 & 0.31 & 0.23 & 0.67 & 1.4 \\
				G002.12+17.45 & 2.122 & 17.445 & 5.56 & 14.9 & 0.5 & 4.2 & 3.7 & 3.9 & 0.31 & 0.23 & 0.67 & 1.4 \\
				G002.15+21.84 & 2.152 & 21.835 & -0.40 & 13.0 & 0.7 & 4.2 & 3.7 & 3.4 & 0.32 & 0.22 & 0.67 & 1.5 \\
				G002.21+17.63 & 2.210 & 17.631 & 5.76 & 18.7 & 0.4 & 5.6 & 5.0 & 4.4 & 0.31 & 0.26 & 0.70 & 1.2 \\
				G002.49+09.75 & 2.487 & 9.750 & 2.89 & 16.5 & 0.5 & 6.2 & 5.5 & 4.3 & 0.41 & 0.24 & 0.82 & 1.7 \\
				G002.50-00.68 & 2.499 & -0.680 & 16.28 & 5.6 & 0.2 & 0.6 & 0.6 & 11.0 & 1.33 & 0.14 & 2.31 & 9.4 \\
				G002.50-00.68 & 2.499 & -0.680 & 8.31 & 8.5 & 0.3 & 5.3 & 4.7 & 4.5 & 2.81 & 0.17 & 4.87 & 16.2 \\
				G002.50-00.68 & 2.499 & -0.680 & 0.17 & 7.8 & 0.6 & 8.6 & 7.7 & 3.2 & 2.35 & 0.17 & 4.08 & 14.1 \\
				G002.55+09.69 & 2.549 & 9.689 & 2.64 & 16.0 & 0.6 & 5.8 & 5.2 & 4.7 & 0.35 & 0.24 & 0.73 & 1.5 \\
				G002.72+09.54 & 2.725 & 9.537 & 2.46 & 12.4 & 0.4 & 3.3 & 2.9 & 5.5 & 0.50 & 0.21 & 0.94 & 2.4 \\
				G002.76+09.43 & 2.764 & 9.429 & 2.46 & 22.2 & 0.4 & 12.2 & 10.8 & 4.6 & 0.58 & 0.28 & 1.12 & 2.1 \\
				G002.79+20.02 & 2.789 & 20.020 & 0.55 & 15.9 & 0.2 & 1.4 & 1.2 & 7.2 & 0.26 & 0.24 & 0.62 & 1.1 \\
				G002.80+07.03 & 2.803 & 7.034 & 5.90 & 9.4 & 0.1 & 0.6 & 0.5 & 9.8 & 0.41 & 0.18 & 0.78 & 2.2 \\
				G002.80+07.03 & 2.803 & 7.034 & 3.00 & 21.0 & 0.3 & 5.3 & 4.7 & 4.3 & 0.36 & 0.27 & 0.78 & 1.3 \\
				G002.83+21.92 & 2.828 & 21.917 & 0.16 & 17.8 & 0.3 & 2.8 & 2.5 & 5.8 & 0.28 & 0.25 & 0.65 & 1.1 \\
				G002.93+09.14 & 2.929 & 9.142 & 2.87 & 15.9 & 0.4 & 7.5 & 6.6 & 4.2 & 0.72 & 0.24 & 1.31 & 3.0 \\
				G002.98+08.72 & 2.981 & 8.724 & 2.78 & 10.0 & 0.8 & 2.4 & 2.1 & 3.7 & 0.28 & 0.19 & 0.59 & 1.5 \\
				G002.99+00.12 & 2.990 & 0.124 & 20.79 & 6.6 & 0.6 & 4.1 & 3.7 & 2.7 & 1.57 & 0.15 & 2.74 & 10.2 \\
				G002.99+00.12 & 2.990 & 0.124 & 12.12 & 7.4 & 0.3 & 4.1 & 3.7 & 4.3 & 2.18 & 0.16 & 3.79 & 13.4 \\
				G002.99+00.12 & 2.990 & 0.124 & 2.98 & 9.5 & 0.5 & 5.4 & 4.8 & 3.7 & 1.09 & 0.18 & 1.92 & 5.9 \\		
				\bottomrule
			\end{tabular}
			\begin{tablenotes}      		    			
				\item [†]This table is available in its entirety in machine-readable form in the online journal. A portion is shown here for guidance regarding its form and content. All observed parameters have been corrected for the main beam efficiency of the telescope.   			
			\end{tablenotes}       	
		\end{threeparttable}
	\end{table*}
	
	\begin{sidewaystable}
		\centering 
		\small
		\setlength{\tabcolsep}{2.5pt}
		\caption{A statistical analysis of parameters from PGCCs}
		\label{tab:3}
		
		\begin{threeparttable}  \begin{tabular}{l *{13}{c}}          
				\toprule
				Parameter & $T_\mathrm{mb}$($^{12}$CO) & FWHM($^{12}$CO) & $T_\mathrm{mb}$($^{13}$CO) & FWHM($^{13}$CO) & $T_\mathrm{ex}$ & $\tau_\mathrm{^{13}CO}$ & $N_\mathrm{^{13}CO}$ & $N_\mathrm{H_2}$ & $\sigma_\mathrm{NT}$ & $c_{s}$ & $\sigma_\mathrm{3D}$ & $I_{^{12}\text{CO}}$/$I_{^{13}\text{CO}}$ & $\mathcal{M}_{\rm ^{13}CO}$ \\
				& (K) & (\kms) & (K) & (\kms) & (K) & & ($10^{15}\mathrm{cm}^{-2}$) & ($10^{21}\mathrm{cm}^{-2}$) & (\kms) & (\kms) & (\kms) & & \\
				\midrule
				\multicolumn{13}{c}{Statistics} \\
				\midrule
				Number  & 2291 & 2291 & 2291 & 2291 & 2291 & 2273 & 2273   & 2273 & 2291  &2291  & 2291  & 2291 & 2291 \\
				Min     & 1.2  & 0.5  & 0.1  & 0.3  & 4.2  & 0.1  &  0.1   & 0.1  & 0.1   &0.1  & 0.4   & 0.7  & 0.6 \\
				Max     & 26.7 & 21.1 & 12.0 & 19.0 & 30.6 & 4.6  &  96.1  & 85.6 & 8.1   &0.3  & 14.0  & 23.6 & 47.6 \\
				Mean    & 7.1  & 3.0  & 2.6  & 2.0  & 10.4 & 0.5  &  4.8   & 4.3  & 0.9   &0.2  & 1.5   & 5.2  & 4.7 \\
				Median  & 6.3  & 2.5  & 2.2  & 1.6  & 9.7  & 0.4  &  3.1   & 2.7  & 0.7   &0.2  & 1.2   & 4.6  & 3.7 \\
				Std dev & 3.4  & 2.1  & 1.8  & 1.6  & 3.5  & 0.3  &  6.1   & 5.5  & 0.7   &0.03  & 1.2   & 2.4  & 4.1 \\
				MAD     & 1.9  & 0.9  & 1.1  & 0.6  & 1.9  & 0.2  &  1.9   & 1.7  & 0.3   &0.02  & 0.4   & 1.2  & 1.5 \\
				\midrule
				\multicolumn{13}{c}{lognormal fit} \\
				\midrule
				$\mu$    & 1.9 & 0.9 & 0.7 & 0.5 & 2.3 & -0.9  &  1.1   & 1.0   & -0.4  &-1.7  & 0.2  & 1.6 & 1.3 \\
				$\sigma$ & 0.5 & 0.6 & 0.7 & 0.6 & 0.3 & 0.6   &  1.0   & 1.0   & 0.6   &0.2  & 0.6  & 0.4 & 0.7 \\
				P        & 0.8 & 0.5 & 0.0 & 0.0 & 0.0 & 0.8   &  0.0   & 0.0   & 0.0   &0.0  & 0.0  & 0.0  & 0.0 \\
				\bottomrule
			\end{tabular}			
			\begin{tablenotes}
				\footnotesize
				\centering
				\item [†]Only parameters involving the $^{13}\mathrm{CO}$ component were used for analysis and calculations, MAD represents the median absolute deviation from the median.
			\end{tablenotes}
		\end{threeparttable}         
	\end{sidewaystable}

\clearpage
\bibliography{ref2}
\bibliographystyle{aasjournalv7}

\end{document}